\documentclass[11pt]{article}
\usepackage{amsmath}
\usepackage{amsfonts}
\usepackage{amscd}
\usepackage{amssymb}
\usepackage{array}
\def\({\left(}
\def\){\right)}

\def\IG{\relax{\rm I\kern-.18em \Gamma}}
\def\be{\begin{equation}}
\def\ee{\end{equation}}
\def\bea{\begin{eqnarray}}
\def\eea{\end{eqnarray}}

\def\bfone{\relax{\rm 1\kern-.35em 1}}

\begin{document}
\begin{titlepage}

\thispagestyle{empty} \vspace{35pt}

\begin{center}{ \LARGE{\bf
Dual Gauged Supergravities}} \vspace{60pt}

{\bf  M. Trigiante$^\bigstar $}

\vspace{15pt}
$^\bigstar${\it Dipartimento di Fisica, Politecnico di Torino \\
C.so Duca degli Abruzzi, 24, I-10129 Torino, and\\
Istituto Nazionale di Fisica Nucleare, \\
Sezione di Torino,
Italy}\\[1mm] {E-mail:  mario.trigiante@polito.it}
\vskip 1,5 cm
\begin{center}
{\it To appear in the Proceedings of the XVII Sigrav Conference,
4-7 September 2006, Turin.}
\end{center}

\vskip 1.5cm
\end{center}
\begin{abstract}
 We shall review a novel formulation of four dimensional
gauged supergravity which is manifestly covariant with respect to
the non-perturbative electric-magnetic duality symmetry
transformations of the ungauged theory,  at the level of the
equations of motion and Bianchi identities. We shall also discuss
the application of this formalism to the description of M-theory
compactified on a twisted torus in the presence of fluxes and to the interpretation from a M/Type IIA
theory perspective of the $D=5\rightarrow D=4$ generalized Scherk-Schwarz reduction. This latter analysis will bring up the issue of
non-geometric fluxes.
\end{abstract}
\end{titlepage}

\section{Introduction}
There are reasons to believe that  superstring theory  in ten
dimensions or the dual M-theory in eleven dimensions may have an
important role in the definition of the fundamental quantum theory
of gravity. Since we live in a four dimensional universe, the
first requirement for any predictable model, based on these
theories, is to encode a mechanism of dimensional reduction from
ten or eleven dimensions to four. The simplest mechanism of this
type is ordinary Kaluza-Klein compactification of string/M--theory
on solutions with geometry of the form $M^{(1,3)}\times
\mathcal{M}$, where $M^{(1,3)}$ is a maximally symmetric four
dimensional space--time with Lorentzian signature and
$\mathcal{M}$ is a compact internal manifold. It is known that the
low--energy dynamics of string/M--theory realized on these
backgrounds, which involve only the massless modes on $M^{(1,3)}$,
is captured (in some cases only in part) by a four dimensional
supergravity theory. In our discussion we shall focus on
compactifications which yield theories in four dimensions with
$N\ge 2$ supersymmetries (\emph{extended} supergravities) on a
Minkowski space-time. Supergravity models on four dimensional
Minkowski vacua, obtained through ordinary Kaluza--Klein reduction
on a Ricci-flat manifold(for instance superstring theory
compactified on a Calabi--Yau 3--fold), are far from being
phenomenologically interesting, since they are typically plagued,
at the classical level, by a plethora of massless scalar fields
$\phi$ (associated for instance with the geometric moduli of the
internal manifold which describe its shape and size) whose vacuum
expectation values define a continuum of degenerate vacua. In fact
there is no dynamics, encoded in some effective scalar potential
$V(\phi)$, which can lift this degeneracy and thus, apart from
predicting massless scalars which are not observed in our real
world, these models also suffer from an intrinsic lack of
predictiveness. An other feature of these supergravities is the
absence of a local internal symmetry gauged by the vector fields.
In other words the vector fields are not minimally coupled to any
other field in the theory. For this reason these models  are also
called \emph{ungauged}.\par Realistic string/M-theory-based models
in four dimensions need to feature a non trivial scalar potential
which could on the one hand lift the moduli degeneracy, thus
making the theory more predictive, and on the other hand select a
vacuum state for our universe with some interesting physical
properties such as for instance spontaneous supersymmetry
breaking, a hierarchy of scales for gravitational and Standard
Model interactions, a (small) positive cosmological constant
 etc. From a higher dimensional perspective the
presence of \emph{fluxes} (for recent reviews on flux
compactifications see \cite{g}) in the space--time background
seems to do the job by inducing a scalar potential in the four
dimensional theory which can fix part or even all the moduli. By
fluxes  here we mean not just the v.e.v. of higher dimensional
$p$--form field strengths across non--trivial cycles of the
internal manifold (\emph{form
fluxes})\cite{formflux,gargiulo,formflux2}, but also background
quantities associated with the geometry of the manifold itself
(\emph{geometric fluxes}) \cite{ss}-\cite{vz}. Recently the
meaning of fluxes has been extended o include background
quantities associated with a new class of manifolds with no
definite global or even local geometry (\emph{non--geometric
fluxes}) \cite{h,dh,stw,glw2}. From a four dimensional
perspective, the only known way for introducing a non--trivial
scalar potential without explicitly breaking supersymmetry is
through the \emph{gauging} procedure  \cite{dwn}-\cite{dwst3},
which consists in promoting a suitable global symmetry group of
the Lagrangian to local symmetry by introducing minimal couplings
for the vector fields, mass deformations and a scalar potential.
Flux compactifications, as opposed to ordinary Kaluza--Klein
compactifications to Minkowski space-time, typically yield gauged
supergravities in four dimensions and a precise statement can be
made about the correspondence between the internal fluxes and the
local symmetry of the lower--dimensional field theory. In some
cases the effective scalar potential $V(\phi)$, at the classical
level, is non--negative and defines vacua with vanishing
cosmological constant in which supersymmetry is spontaneously
broken and part of the moduli are fixed. Models of this type are
generalizations of the so called ``no--scale'' models
\cite{noscale} which were subject to intense study during the
eighties. Special care is needed in defining a limit in which
(gauged) supergravity is  reliable as a low--energy description of
a flux compactification. In many cases it suffices for the back
reaction of the fluxes on the background geometry to be
negligible. This regime, for instance in the case of form-fluxes,
can be attained if the size of the internal manifold is much
larger than the string scale, so that the following hierarchy of
scales is realized: (flux-induced masses) $\ll$ Kaluza--Klein
masses $\ll$ mass of string excitations. In other cases, the
gauged supergravity only describes a consistent truncation of the
lowest lying modes, as for the compactification of M-theory on a
seven--sphere \cite{dwn}.
\par The recent study of this new kind of string/M--theory
compactifications had opened a Pandora's box containing an
enormous number of possible microscopic settings and thus of
vacua. We are left with the hope that on the one hand this picture
may simplify by the presence of some duality symmetry underlying
the landscape of vacua, as a consequence of which several of these
solutions can be thought of as different descriptions of a single
microscopic one. On the other hand that there could exist some,
yet unknown, dynamic mechanism which can select one vacuum out of
the many.\par The paper is organized as follows. In section 2,
after recalling the main facts about global symmetries in extended
four dimensional supergravities and their relation to string
dualities, we shall introduce the concept of \emph{embedding
tensor} \cite{cgftt,dwst1,dwst2,dws,dwst3}
 in terms of which we shall make a precise statement about the correspondence between flux compactifications and gauged supergravity.
In section 3 we shall give a formal discussion of gauged
supergravities, focusing mainly on the maximally supersymmetric
theory. In section 4 a novel formulation of these models
\cite{dwst3} is reviewed, in which the global symmetries of the
ungauged theory are restored at the level of field equations and
Bianchi identities. In section 5 we shall discuss an application
of this machinery to the specific example of M--theory
compactified on a ``twisted'' torus in the presence of fluxes and
show how two different four dimensional gauged supergravities can
actually be considered as ``dual'' descriptions of the same
compactification. In section 6 we shall identify the components of
the embedding tensor corresponding to the ``non--geometric''
fluxes which are T-dual to the NS-NS 3-form flux. We refer the
reader to appendix A for a detailed, group theoretical discussion
of the flux-embedding tensor correspondence. In section 7 the
parameters of the so called ``generalized Scherk-Schwarz
reduction'' from $D=5$ to $D=4$ \cite{css,adfl} will be described
in terms of M--theory/Type IIA form-, geometric and non-geometric
fluxes. In particular we will show that one of these parameters
can be described in terms of U-dual fluxes. We shall end with some
final comments and outlook.
\section{The role of global symmetries}
An important role in understanding several non--perturbative
aspects of superstring theory has been played by the global
symmetries of the lower dimensional supergravity. Behind the
concept of string duality there is the idea that superstring
theories or M-theory on various backgrounds, are just different
realizations of a unique fundamental quantum theory and the
correspondence among them is called duality. Upon ordinary
dimensional reduction to  four dimensional Minkowski space--time,
these dualities are conjectured to be encoded in the global
symmetries of the resulting (ungauged) supergravity \cite{ht}. A
wide class of ungauged extended supergravities feature at the
classical level a continuous group o global symmetries which act,
as we shall see, as generalized electric--magnetic dualities.
Already at the field theory level, Dirac-Zwanziger quantization
condition on electric and magnetic charges causes this global
symmetry group to break to a suitable discrete subgroup. It is the
latter which is conjectured to describe the string/M--theory
dualities. Here we shall restrict our analysis to classical
supergravity only. There are two important features of these
global symmetries.
\begin{itemize}
\item In four dimensional ungauged supergravity, antisymmetric
tensor fields and scalar fields are related by Poincar\'e duality.
The amount of global symmetry of the theory depends on the number
of antisymmetric tensor fields which have been dualized into
scalar fields. It is maximal, and we shall denote it by $G$, when
all antisymmetric tensors are dualized into scalar fields. This
phenomenon is called \emph{Dualization of Dualities} and was
studied in \cite{dd}.
\item In the presence of internal fluxes, the lower-dimensional supergravity is no longer ungauged
but features minimal couplings and, by consistency of the theory,
local symmetries. Since minimal couplings involve only electric
vector fields the electric-magnetic duality symmetry of the
ungauged theory, obtained in the limit of zero fluxes,  is
manifestly broken by the background quantities.
\end{itemize}
In fact, in all known instances of compactification, fluxes enter
the four dimensional gauged Lagrangian $\mathcal{L}_g$ not just in
the minimal couplings, but they also determine mass terms and a
scalar potential. These background quantities, as it was  shown in
\cite{km} in the heterotic theory, can be associated with
representations of the global symmetry group $G_e\subset G$ of the
ungauged Lagrangian $\mathcal{L}_{u-g}$ so that the original
$G_e$--invariance is restored at the level of $\mathcal{L}_g$ if
the fluxes are transformed under $G_e$ as well. It turns out that
we can make an even stronger statement:\par \emph{Fluxes can be
associated with representations of the whole global symmetry group
$G$, so that, if transformed accordingly,  the original
$G$--invariance  is restored at the level of the gauged equations
of motion and the Bianchi identities.}\par However $G$ can no
longer be regarded as a symmetry of the gauged theory, since it
has a non-trivial action on the background quantities (coupling
constants), but rather it should be thought of as a mapping
(duality) between different theories, i.e. different
compactifications.\par
 How can fluxes be associated with representations of the full electric-magnetic duality group $G$?
 It was found in \cite{dwst1} that the most general gauge group which can be
 introduced in an extended supergravity can be described in terms
 of a $G$--covariant tensor called the \emph{embedding tensor}. It
 turns out that in all known instances of flux compactifications,
 fluxes enter the lower dimensional gauged supergravity as
 components of the embedding tensor:
\begin{eqnarray}
\mbox{internal flux}&\leftrightarrow & \mbox{embedding tensor
$\Theta$}\,.
\end{eqnarray}
  This identification represents a precise statement on the
  correspondence between flux compactifications and local symmetry
  of the lower--dimensional supergravity.\par
  Since $G$ acts, in extended supergravities, as a generalized
  electric-magnetic duality, in order to achieve full $G$
  covariance of the gauged field equations and Bianchi identities,
  it is necessary to introduce magnetic couplings besides the electric ones. As we shall see
  this can be done in a local field theory provided the embedding
  tensor satisfies certain locality conditions and moreover it
  requires the introduction in the theory of antisymmetric tensor
  fields. We shall illustrate this new construction of gauged
  extended supergravities in the maximal case, namely for the
  $N=8$ theory in four dimensions. The ungauged version of this theory, with no antisymmetric tensor
  fields, was constructed in \cite{cj} by dimensionally reducing eleven dimensional supergravity \cite{cjs} (which describes the low-energy limit of M-theory)
 on a seven torus $T^7$ and then dualizing seven antisymmetric tensors to scalar fields. This theory was shown to exhibit a duality symmetry $G={\rm
  E}_{7(7)}$ at the level of field equations and Bianchi
  identities.
\section{The $N=8,\,D=4$ supergravity}
The four dimensional maximal supergravity is characterized by
having $N=8$ supersymmetry (that is $32$ supercharges), which is
the maximal amount of supersymmetry allowed in order for the
theory to be local. As anticipated we shall restrict ourselves to
the ungauged $N=8$ theory with no antisymmetric tensor field. The
theory describes a single massless graviton supermultiplet
consisting of the graviton $g_{\mu\nu}$, eight spin $3/2$
gravitini $\psi^A_\mu$ ($A=1,\dots, 8$) transforming in the
fundamental representation of the R--symmetry group ${\rm SU}(8)$,
$28$ vector fields $A^\Lambda_\mu$, $\Lambda=0,\dots, 27$, $56$
spin $1/2$ ``dilatini'' $\chi_{ABC}$ transforming in the ${\bf
56}$ of ${\rm SU}(8)$ and finally $70$ real scalar fields
$\phi^r$. A common feature of supergravity theories is that the
scalar fields are described by a non--linear $\sigma$--model on a
target space which is a Riemannian manifold $\mathcal{M}_{scal}$.
In other words the scalar fields are local coordinates on
$\mathcal{M}_{scal}$ and the scalar action is invariant under the
global action of the isometry group ${\rm
Isom}(\mathcal{M}_{scal})$ of $\mathcal{M}_{scal}$ on the scalars.
For $N>2$ supersymmetry requires  $\mathcal{M}_{scal}$ to be a
homogeneous symmetric manifold, namely a manifold of the form
$G/H$ where the isometry group  $G={\rm Isom}(\mathcal{M}_{scal})$
is a semisimple Lie group and $H\subset G$ its maximal compact
subgroup. In the $N=8$ model, the scalar manifold has the form
\begin{eqnarray}
\mathcal{M}_{scal}&=&\frac{G}{H}=\frac{{\rm E}_{7(7)}}{{\rm
SU}(8)}\,,
\end{eqnarray}
the isometry group  being $G={\rm E}_{7(7)}$ and $H={\rm SU}(8)$
is the R--symmetry group. Although we shall mainly be interested
in the maximal theory, most of our treatment will hold also for
extended supergravities with lower supersymmetry, describing $n_v$
vector fields and scalar fields $\phi^r$ spanning a homogeneous
manifold of the form $\mathcal{M}_{scal}=G/H$. For non-maximal
theories the subgroup $H$ will have the general form
$H=H_{R}\times H_{matter}$, where $H_{R}$ is the R--symmetry group
and $H_{matter}$ is a compact group acting on the matter fields.
The gravitino and fermion fields will transform in representations
of $H$.
\par The bosonic action of an ungauged supergravity model has the
following general form
\begin{eqnarray}
 {\cal S}_{u-g}&=&\int \mathcal{L}_{u-g}=\int
d^4x\left(-\frac{e}{2}\,R+\frac{e}{4}\,{\rm Im}{\cal N}_{\Lambda
\Gamma}F_{\mu\nu}^{\Lambda } F^{\Gamma |\mu\nu}+ \frac{1}{8}\,{\rm
Re}{\cal N}_{\Lambda \Gamma  } \epsilon^{\mu\nu\rho\sigma}\,
F_{\mu\nu}^{\Lambda } F^{\Gamma
}_{\rho\sigma}+\right.\nonumber\\
&+&\left.\frac{e}{2} \,g_{rs}(\phi) \partial_{\mu}
\phi^{r}\partial^{\mu}\phi^{s}\right)\,,\label{bosonicL}
\end{eqnarray}
where $F_{\mu\nu}^{\Lambda }=2\,\partial_{[\mu}A^\Lambda_{\nu]}$
 ($\Lambda=0,\dots, n_v-1$) are the vector field strengths and $g_{rs}(\phi)$ is the metric on
the scalar manifold. We can associate with the electric field
strengths $F_{\mu\nu}^{\Lambda }$ their magnetic ``duals''
$G_{\mu\nu\,\Lambda}$ defined as follows
\begin{eqnarray}
{{}^\star G}_{\Lambda|\mu\nu} \, \equiv \, 2\,{ \frac{\partial
{\cal L}}{\partial F^\Lambda_{\mu\nu}}}\,,\label{defG}
\end{eqnarray}
where ${}^\star$ denotes the Hodge duality operation: ${}^\star
F_{\mu\nu}=\frac{e}{2}\,\epsilon_{\mu\nu\rho\sigma}\,F^{\rho\sigma}$.
In terms of $F^\Lambda$ and $G_{\Lambda}$ the Maxwell equations
read
\begin{eqnarray}
\nabla^{\mu }{{}^\star  F}^{\Lambda}_{\mu\nu} &=& 0\,\,\,;\,\,\,\,\nabla^{\mu }{{}^\star G}_{\Lambda |\mu\nu} = 0\,.
\label{biafieq}
\end{eqnarray}
 A general feature of
(\ref{bosonicL}) is that the scalar fields enter the kinetic terms
of the vector fields through the complex symmetric matrix
$\mathcal{N}_{\Lambda\Sigma}(\phi)$ whose negative definite
imaginary part generalizes the inverse of the squared coupling
constant appearing in ordinary gauge theories while its real part
is instead a generalization of the \emph{theta}-angle of quantum
chromodynamics. As a consequence of this, a symmetry
transformation of the scalar field part of the action, will not in
general leave the vector field part invariant. In extended
supergravities however, as it was shown in \cite{gz}, the global
symmetry group $G$ of the scalar action can be promoted to global
invariance of at least the field equations and the Bianchi
identities, provided its (non--linear) action on the scalar fields
can be associated with a linear transformation on the vector field
strengths $F^\Lambda_{\mu\nu}$ and their magnetic duals
$G_{\mu\nu\,\Lambda}$ \footnote{In most cases this linear action
can be associated with all the isometries of the scalar manifold.
However this is not a general rule. We are grateful to A. Van
Proeyen for making this point.}:
\begin{eqnarray}
 g\in G &: &\begin{cases}\,\,\,\,\,\,\phi^r &\rightarrow \,\,\phi^r_g(\phi) \,\,\,\,\qquad\qquad\qquad\qquad\qquad\qquad\qquad\mbox{(non--linear action)}\cr
 \left(\begin{matrix}F^\Lambda\cr G_\Lambda\end{matrix}\right)&\rightarrow \,\,\iota(g)\cdot\left(\begin{matrix}F^\Lambda\cr G_\Lambda\end{matrix}\right)=
 \left(\begin{matrix}A^\Lambda{}_\Sigma&
 B^{\Lambda\Sigma}\cr C_{\Lambda\Sigma}& D_\Lambda{}^\Sigma\end{matrix}\right)\,\left(\begin{matrix}F^\Sigma\cr G_\Sigma\end{matrix}\right)\mbox{(linear
 transformation)}\end{cases}\,.\nonumber\\
\end{eqnarray}
The linear transformation $\iota(g)$ associated with the element
$g$ of $G$ mixes electric and magnetic field strengths and
therefore acts as a generalized electric--magnetic duality.
Consistency of this transformation with the definition
(\ref{defG}) of $G_\Lambda$ requires $\iota(g)$ to be a symplectic
transformation, namely it should leave the following $2n_v\times
2n_v$ antisymmetric matrix
\begin{eqnarray}
\Omega&=&\left(\begin{matrix}{\bf 0}_{n_v}&
 \bfone_{n_v}\cr -\bfone_{n_v}& {\bf
 0}_{n_v}\end{matrix}\right)\,,\label{Omega}
\end{eqnarray}
invariant: $ \iota(g)^T\cdot\Omega\cdot \iota(g)=\Omega\,.$
Summarizing, if $n_v$ is the number of vector fields of the model
($n_v=28$ in the maximal case) then the global symmetry group of
the field equations and Bianchi identities acts as a duality
transformation defined by the embedding $\iota$ of the isometry
group $G$ of the scalar manifold into the symplectic group ${\rm
Sp}(2\,n_v,\,\mathbb{R})$
\begin{eqnarray}
G&\stackrel{\iota}{\mbox{{\LARGE $\hookrightarrow$}}}& {\rm
Sp}(2\,n_v,\,\mathbb{R})\,.
\end{eqnarray}
As a consequence of this, the electric field strengths and their
magnetic duals transform in the ${\bf 2\,n_v}$ symplectic
representation of $G$. In the maximal case $G={\rm E}_{7(7)}$ and
 the electric and magnetic charges fill the ${\bf 56}$
representation, which is symplectic. The duality action $\iota(G)$
of $G$ is not unique, since it depends on which elements of the
basis of the ${\bf 2\,n_v}$ representation are chosen to be the
$n_v$ elementary vector fields, to be described locally in the
Lagrangian, and which their magnetic duals. This amounts to
choosing the \emph{symplectic frame} and determines the embedding
of $G$ inside ${\rm Sp}(2\,n_v,\,\mathbb{R})$, which is not
unique. Different choices of the symplectic frame may yield
inequivalent Lagrangians (namely Lagrangians not related by local
field redefinitions) with different global symmetries. Indeed the
global symmetry group of the Lagrangian is defined as the subgroup
$G_e$ of $G$ whose duality action is linear on the electric field
strengths
\begin{eqnarray}
\iota(G_e)&=&\left(\begin{matrix}A^\Lambda{}_\Sigma&
 {\bf 0}\cr C_{\Lambda\Sigma}&
 D_\Lambda{}^\Sigma\end{matrix}\right)\,\Rightarrow\,\begin{cases}F^\Lambda\rightarrow A^\Lambda{}_\Sigma\,F^\Sigma\cr
 G_\Lambda\rightarrow C_{\Lambda\Sigma}\,F^\Sigma+
 D_\Lambda{}^\Sigma\,G_\Sigma\end{cases}\,.\label{ge}
\end{eqnarray}
\subsection{The gauging}
As anticipated in the introduction, the gauging procedure consists
in promoting a suitable global symmetry group $\mathcal{G}\subset
G_e $ of the Lagrangian to local symmetry gauged by the vector
fields of the theory, and therefore we must have ${\rm dim}(\mathcal{G})\le n_v$. The first
condition for a global symmetry group $\mathcal{G}$ to be a viable
gauge group is that there should exist a subset of the vector
fields $A^\Lambda_\mu$ which transform under the duality action of
$\mathcal{G}$ in its co--adjoint representation. These fields will
become the gauge vectors associated with the generators
$X_\Lambda$ of $\mathcal{G}$. If we denote by $\Omega_\mu$ the
gauge connection, it will have the form
\begin{eqnarray}
\Omega_\mu&=&A^\Lambda_\mu\,X_\Lambda\,.
\end{eqnarray}
The gauge algebra, which need not be compact or even semisimple,
is characterized by the structure constants
$f_{\Lambda\Sigma}{}^\Gamma$ which define the commutation
relations
\begin{eqnarray}
\left[X_\Lambda,\,X_\Sigma\right]&=&f_{\Lambda\Sigma}{}^\Gamma\,X_\Gamma\,.
\end{eqnarray}
Being $\mathcal{G}$ a subgroup of $G_e$, its generators
$X_\Lambda$ will have an electric-magnetic duality action which is
represented by a symplectic matrix of the form (\ref{ge})
\begin{eqnarray}
X_\Lambda&\equiv&\left(\begin{matrix}X_\Lambda{}^\Sigma{}_\Gamma&
 {\bf 0}\cr X_{\Lambda\Sigma\Gamma}&
 X_{\Lambda\Sigma}{}^\Gamma\end{matrix}\right)\,.\label{xsym}
\end{eqnarray}
The symplectic condition on the matrix form of $X_\Lambda$ implies
the following properties:
$X_\Lambda{}^\Sigma{}_\Gamma=-X_{\Lambda\Gamma}{}^\Sigma$,
$X_{\Lambda\Sigma\Gamma}=X_{\Lambda\Gamma\Sigma}$.
 The first step in the gauging procedure is to associate
the fields with representations of $\mathcal{G}$ and to
covariantize the derivatives acting on them accordingly (thus
introducing the minimal couplings)
\begin{eqnarray}
\partial_\mu & \longrightarrow & \nabla_\mu=\partial_\mu
-g\,A_\mu^\Lambda\,X_\Lambda\,,
\end{eqnarray}
$g$ being the coupling constant. If the symplectic duality action
(\ref{xsym}) of $X_\Lambda$ has a non--vanishing off diagonal
block $X_{\Lambda\Sigma\Gamma}$, gauge invariance further requires
the addition  to the Labrangian of a topological term \cite{dwlv}
of the form
\begin{eqnarray}
\mathcal{L}_{top}&=&-\frac{1}{3}\,g\,\epsilon^{\mu\nu\rho\sigma}\,X_{\Lambda\Sigma\Gamma}\,A_\mu^\Lambda\,A_\nu^\Sigma\,
\left(\partial_\rho
A^\Gamma_\sigma+\frac{3}{8}\,g\,X_{\Xi\Omega}{}^\Gamma\,A_\mu^\Xi\,A_\nu^\Omega\right)\,,\label{top}
\end{eqnarray}
provided the following condition holds
\begin{eqnarray}X_{(\Lambda\Sigma\Gamma)}=0\,,\label{xsymm}\end{eqnarray}
where  total symmetrization of the indices within the round
brackets is understood. As we shall see below, condition
(\ref{xsymm}) is a consequence of the constraints on the gauge
algebra which are required by supersymmetry. Indeed, although the
resulting Lagrangian will be locally $\mathcal{G}$--invariant, the
minimal couplings will explicitly break both supersymmetry and the
duality symmetry $G$. In order to restore supersymmetry one needs
to further deform the Lagrangian in the following way.
\begin{itemize}\item Add order--$g$ terms to the  supersymmetry transformation rules
for the gravitino ($\psi_{\mu A}$) and the fermion fields ($\chi^\mathcal{I}$), which are
characterized by some scalar-dependent matrices $S_{AB},\, N^{\mathcal{I} A}$,
called the \emph{fermion shift matrices}, in appropriate
representations of the $H$--group ($A,\,B$ are the indices labelling the supersymmetry generators and $\mathcal{I},\mathcal{J}$
are the generic indices labelling the fermion fields)
\begin{eqnarray}
\delta_\epsilon\psi_{\mu A}&=&g\,S_{AB}\,\gamma_\mu\,\epsilon^B+\dots\,,\nonumber\\
\delta_\epsilon \chi^{\mathcal{I}}&=&g\,N^{\mathcal{I}A}\,\epsilon_A+\,\epsilon\,.
\end{eqnarray}
\item Add gravitino and fermion mass terms to the Lagrangian
defined by the shift matrices
\begin{eqnarray}
e^{-1}\,\mathcal{L}&=&\dots+
g\,\bar{\psi}^A_\mu\,\gamma^{\mu\nu}\,S_{AB}\,\psi_\nu^B+g\,\bar{ \chi}_{\mathcal{I}}\,\gamma^\mu\,N^{\mathcal{I}A}\psi_{\mu A}\,.
\end{eqnarray}
\item Finally add an order $g^2$ scalar potential $V(\phi)$ whose expression
is totally fixed as a bilinear in the shift matrices by
supersymmetry
\begin{eqnarray}
\delta^A_B\,V(\phi)&\sim & g^2\,(N^{\mathcal{I}A}\,N_{\mathcal{I}B}-3\,S^{AC}\,S_{BC})\,.
\end{eqnarray}
\item Not for all choices of the gauge group supersymmetry can be restored by the above
prescription. There are further constraints on the Lie algebra of
$\mathcal{G}$ (\emph{supersymmetry constraints}) which need to be
satisfied. These constraints are linear and quadratic in the gauge
generators $X_\Lambda$ and we shall discuss them below in a
general context.
\end{itemize}
It is useful to encode all the information about the gauge algebra
in a single $G_e$--covariant tensor $\theta_\Lambda{}^\sigma$
($\Lambda=1,\dots, n_v$, $\sigma=1,\dots, {\rm dim}(G_e)$), called
the embedding tensor, which expresses the gauge generators as
linear combinations of the global symmetry generators $t_\sigma$
of $G_e$
\begin{eqnarray}
X_\Lambda&=&\theta_\Lambda{}^\sigma\,t_\sigma\,\,\,;\,\,\,\,\theta_\Lambda{}^\sigma\in
{\bf n_v}\times {\rm Adj}(G_e)\,.
\end{eqnarray}
The advantage of this description is that the $G_e$
invariance of the original ungauged Lagrangian $\mathcal{L}_{u-g}$
is restored at the level of the gauged Lagrangian
$\mathcal{L}_{g}$ provided $\theta_\Lambda{}^\sigma$ is
transformed under $G_e$ as well. However the full global symmetry group $G$
of the field equations and Bianchi identities is still broken
since the parameters $\theta_\Lambda{}^\sigma$ can be viewed as
${\rm dim}(G_e)$ electric charges whose presence manifestly break
electric-magnetic duality invariance.\par In order to restore the
original  $G$--invariance, we would need to introduce magnetic
components of the embedding tensor as well. The natural way of
doing this is by extending the definition of $\theta$ to a
$G$--covariant tensor
\begin{eqnarray}
\theta_\Lambda{}^\sigma &\longrightarrow &
\theta_n{}^\alpha\equiv(\theta^{\Lambda\,\alpha},\,\theta_\Lambda{}^\alpha)\,\in\,{\bf
2\,n_v}\times {\rm Adj}(G)\,,
\end{eqnarray}
where the index $n=1,\dots, 2\,n_v$ labels the ${\bf 2\,n_v}$
representation of $G$ (the ${\bf 56}$ representation of ${\rm
E}_{7(7)}$ in the maximally supersymmetric case) and we have
expressed a generic vector in this representation by
$W^n=(W^\Lambda,\,W_\Lambda)$,  $\alpha=1,\dots, {\rm dim}(G)$
labels the generators $t_\alpha$ of $G$. In the maximally
supersymmetric case, for example, this generalized embedding
tensor is associated with the ${\bf 56}\times{\bf 133}$
representation of ${\rm E}_{7(7)}$. Consistency of this definition
requires that ${\rm rank}\,(\theta)\le n_v$ since no more than the
available vector fields can be involved in the minimal couplings.
The linear supersymmetry constraint amounts to a condition on the
$G$--representation of $\theta$. For instance, in the maximally
supersymmetric case it requires $\theta$ to transform in the ${\bf
912}$ representation of ${\rm E}_{7(7)}$ contained in the
decomposition of the ${\bf 56}\times{\bf 133}$. This condition has
been solved explicitly in \cite{dwst1}. The quadratic constraint,
on the other hand, can be viewed as the condition that $\theta$ be
invariant under the action of the gauge group itself.\par Let us
now write the $G$ generators $t_\alpha$, in the ${\bf 2\,n_v}$
representation, as $2n_v\times 2n_v$ matrices $(t_{\alpha})_n{}^m$.
These matrices belong to the algebra of ${\rm
Sp}(2\,n_v,\,\mathbb{R})$ and therefore they satisfy the following property
\begin{eqnarray}
(t_{\alpha})_n{}^m\,\Omega_{mp}&=&(t_{\alpha})_p{}^m\,\Omega_{mn}\,,\nonumber
\end{eqnarray}
where the symplectic invariant matrix $\Omega=\{\Omega_{mn}\}$ was
defined in (\ref{Omega}). It is convenient for what follows to
introduce the following $G$--tensor
$X_{mn}{}^p=\theta_m{}^\alpha\,(t_{\alpha})_n{}^p$. This tensor
can be viewed as the matrix representation in the ${\bf 2\,n_v}$
of $2\,n_v$ gauge generators $X_m\equiv\{(X_m)_n{}^p\}$. Of course
this is just a symplectic covariant notation, and, as we shall see below, the properties
of $\theta$ guarantee that only at most $n_v$ out of the $X_n$
generators are actually independent. However if the embedding
tensor has magnetic components, the symplectic representation of
$X_n$ will also feature a non-vanishing upper off diagonal block $ X_{n}{}^{\Sigma\Gamma}$, besides the blocks
$X_n{}^\Sigma{}_\Gamma$, $ X_{n\Sigma\Gamma}$ and $X_{n\Sigma}{}^\Gamma$.
It was shown in \cite{dwst2} that the linear and quadratic
supersymmetry constraints can be recast in the following
$G$--covariant form:
\begin{eqnarray}
\mbox{Linear constraint:}&& X_{(mnp)}=0\,,\label{linear}\\
\mbox{Quadratic constraint:}&&
X_{mn}{}^p\,X_{qp}{}^\ell-X_{qn}{}^p\,X_{mp}{}^\ell+X_{mq}{}^p\,X_{pn}{}^\ell=0\,,\label{quadratic}
\end{eqnarray}
where $X_{mnp}\equiv X_{mn}{}^q\,\Omega_{qp}$. Note that equation
(\ref{linear}) implies condition (\ref{xsymm}). The $X_{mn}{}^p$
tensor provides a symplectic covariant description of the structure
constants. Indeed equations (\ref{quadratic}) can also be written
as a $G$--covariant closure condition of the gauge algebra
\begin{eqnarray}
\left[X_m,\,X_n\right]&=&-X_{mn}{}^p\,X_p\,.
\end{eqnarray}
The supersymmetry constraints (\ref{linear}) and (\ref{quadratic})
also imply the following quadratic condition on the embedding
tensor
\begin{eqnarray}
\Omega^{mn}\,\theta_m{}^\alpha\,\theta_n{}^\beta=0\,\,\,&
\Leftrightarrow &\,\,\,\theta^{\Lambda
[\alpha}\,\theta_\Lambda{}^{\beta]}=0\,,\label{tad}
\end{eqnarray}
which ensures that ${\rm rank}\,(\theta)\le n_v$, namely that no
more than $n_v$ vector fields will be involved in the minimal
couplings and, as we shall see, guarantees also locality of the
resulting theory with magnetic couplings. In fact, in virtue of
eq. (\ref{tad}), $\theta_n{}^\alpha$ can always be brought, by
means of a symplectic transformation $E_n{}^m$, into an
``electric'' form in which all the magnetic components are zero
\begin{eqnarray}
(\theta^{\Lambda\,\alpha},\,\theta_{\Lambda}{}^\alpha)&\stackrel{E}{\longrightarrow}&
(0,\,\theta^\prime_{\Lambda}{}^\alpha)\,.\label{E}
\end{eqnarray}
In this symplectic frame  the gauged Lagrangian therefore features
electric couplings only.\par
 In \cite{dwst2} this formalism was applied to the study of Type IIB
toroidal compactification to four dimensions in the presence of
internal RR and NS-NS fluxes: $F^{(3)},\,H^{(3)}$. These
background quantities were consistently identified with components
of $\theta$ and the condition (\ref{tad}) was equivalent to the
tadpole cancellation condition $F^{(3)}\wedge H^{(3)}=0$ in the
absence of localized sources.
\section{Duality covariant gauged supergravities}
In \cite{dwst3} a formulation of gauged extended supergravity was
given in which the Lagrangian $\mathcal{L}_g$ accommodates both
the ``electric'' ($\theta_\Lambda{}^\alpha$) and ``magnetic''
($\theta^{\Lambda\,\alpha}$) charges coupled in a symplectic
invariant way to electric and magnetic gauge fields. The
ingredients for this construction are:
\begin{itemize}
\item Antisymmetric tensor fields $B_{\mu\nu\,\alpha}$
transforming in the adjoint representation of $G$ (133 tensor
fields in the maximally supersymmetric theory). The presence of
these fields is related to the magnetic components of the
embedding tensor since they enter the action only in the
combinations $\theta^{\Lambda\,\alpha}\,B_{\mu\nu\,\alpha}$.
Therefore in the ``electric'' symplectic frame in which
$\theta^{\Lambda\,\alpha}=0$ the antisymmetric tensor fields will
disappear all together and we are back to the standard formulation
of gauged supergravity;
\item $n_v$ magnetic vector fields $A_{\mu\,\Lambda}$ which,
together with the existing electric ones $A^\Lambda_\mu$, define
$2\,n_v$ vector fields $A^n_\mu$ transforming in the ${\bf
2\,n_v}$ of $G$;
\item Additional tensor and vector gauge invariance which
guarantees the correct counting of degrees of freedom.
\end{itemize}
This result generalizes previous work in \cite{previous}. Let us
review the main features of this model. The electric and magnetic
gauge fields enter the gauge connection in a symplectic invariant
fashion
\begin{eqnarray}
\Omega_\mu&=&g\,A^n_\mu\,X_n=g\,A^\Lambda_\mu
X_\Lambda+g\,A_{\mu\,\Lambda} X^\Lambda\,,
\end{eqnarray}
and the corresponding gauge field strengths are defined by the
following $G$--covariant expression
\begin{eqnarray}
F^n_{\mu\nu}&=& \partial_\mu A^n_\nu-\partial_\nu A^n_\mu +
g\,X_{[m\ell]}{}^n\,A^m_\mu\,A^\ell_\mu\,.
\end{eqnarray}
The gauge covariant quantities however are not $F^n_{\mu\nu}$ but
rather the following combinations
$H^n_{\mu\nu}=(H^\Lambda_{\mu\nu},\,H_{\mu\nu\,\Lambda})$ of
$F^n_{\mu\nu}$ and the antisymmetric tensors $B_{\mu\nu\,\alpha}$
\begin{eqnarray}
    H^\Lambda_{\mu\nu}&=& F^\Lambda_{\mu\nu}+
\frac{g}{2}\,\theta^{\Lambda \alpha}\,B_{\alpha\,\mu\nu}\,,        \nonumber\\
H_{\Lambda\,\mu\nu}&=& F_{\Lambda\,\mu\nu}-
\frac{g}{2}\,\theta_\Lambda{}^{\alpha}\,B_{\alpha\,\mu\nu}\,.\nonumber
\end{eqnarray}
Off shell the magnetic quantities $H_{\mu\nu\,\Lambda}$ are not
identified with the ``dual'' field strengths $G_{\mu\nu\,\Lambda}$
associated with $ H^\Lambda_{\mu\nu}$: $H_{\mu\nu\,\Lambda}\neq
G_{\mu\nu\,\Lambda}\equiv -e\,\epsilon_{\mu\nu\rho\sigma}\frac{\partial
\mathcal{L}_g}{H^\Lambda_{\rho\sigma}}$.
The bosonic Lagrangian reads \cite{dwst3}
\begin{eqnarray}
{\cal L}_g &=&-\frac{e}{2} R+ \frac{e}{4} \, {\rm Im}(\mathcal{N})
_{\Lambda\Sigma}\,H_{\mu\nu}{}^{\Lambda} H^{\mu\nu\,\Sigma}
+\frac{1}{8}  {\rm
Re}(\mathcal{N})_{\Lambda\Sigma}\;\varepsilon^{\mu\nu\rho\sigma}
H_{\mu\nu}{}^{\Lambda} H_{\rho\sigma}{}^{\Sigma}-
  \nonumber\\[.9ex]
&&{} -\frac{1}{8}\,g\, \varepsilon^{\mu\nu\rho\sigma}\,
\Theta^{\Lambda\alpha}\,B_{\mu\nu\,\alpha} \, \Big(
2\,\partial_{\rho} A_{\sigma\,\Lambda} + g\, X_{mn\,\Lambda}
\,A_\rho{}^m A_\sigma{}^n
-\frac{1}{4}\,g\,\Theta_{\Lambda}{}^{\beta}B_{\rho\sigma\,\beta}
\Big)
\nonumber\\[.9ex]
&&{} -\frac{1}{3}\,g\,
\varepsilon^{\mu\nu\rho\sigma}X_{mn\,\Lambda}\, A_{\mu}{}^{m}
A_{\nu}{}^{n} \Big(\partial_{\rho} A_{\sigma}{}^{\Lambda}
+\frac{1}{4} g\,X_{pq}{}^{\Lambda}
A_{\rho}{}^{p}A_{\sigma}{}^{q}\Big)
\nonumber\\[.9ex]
&&{} -\frac{1}{6}\,g\,
\varepsilon^{\mu\nu\rho\sigma}X_{mn}{}^{\Lambda}\, A_{\mu}{}^{m}
A_{\nu}{}^{n} \Big(\partial_{\rho} A_{\sigma}{}_{\Lambda}
+\frac{1}{4}\, g\,X_{pq\Lambda}
A_{\rho}{}^{p}A_{\sigma}{}^{q}\Big)+\mathcal{L}_{\mbox{matter}}\,,\label{lgn}
\nonumber
\end{eqnarray}
where $\mathcal{L}_{\mbox{matter}}$ is the gauge invariant scalar
action. The topological terms in the above Lagrangian generalize
to the presence of antisymmetric tensors, the Chern-Simons-like
term in (\ref{top}). The Lagrangian (\ref{lgn}) is invariant with
respect to the following vector gauge transformations,
parametrized by $2\,n_v$ local parameters
$\Lambda^n=(\Lambda^\Sigma,\,\Lambda_\Sigma)$, and tensor gauge
transformations, parametrized by 1--forms $\Xi_{\mu\,\alpha}$,
\begin{eqnarray}
\delta A^n_\mu&=&D_\mu \Lambda^n-\frac{g}{2}\,
\theta^{n,\alpha}\,\Xi_\alpha\,,
\nonumber\\
\theta^{\Lambda\alpha}\delta B_{\mu\nu\alpha}&=&2\,
\theta^{\Lambda\alpha}\left[D_{[\mu}\Xi_{\nu]\alpha} +t_{\alpha
mn}\, A^m_{[\mu} \delta A^n_{\nu]}\right]-2
\Lambda^n\left[X^\Lambda{}_{n\Sigma}
\,H^\Sigma_{\mu\nu}-X^\Lambda{}_{n}{}^{\Sigma}
\,G_{\Sigma\mu\nu}\right]\,,\label{gauge}
\end{eqnarray}
where $\theta^{n\alpha}=\Omega^{nm}\,\theta_{m}{}^\alpha$ and
$D_\mu\Lambda^n=\partial_\mu\Lambda^n+g\,X_{mp}{}^n\,A^m_\mu\,\Lambda^p$.\par
The field equations obtained by varying the Lagrangian with
respect to the tensor fields and the vector fields have the
following $G$--covarinat form
\begin{eqnarray}
\frac{\delta \mathcal{L}_g}{\delta
B_{\mu\nu\alpha}}=0\,\,\,\,&\Leftrightarrow&\qquad\qquad
\theta_n{}^\alpha\,(G_{\mu\nu}{}^n-H_{\mu\nu}{}^n)=0\,,\qquad\qquad\qquad\qquad\label{deltab}\\
\frac{\delta \mathcal{L}_g}{\delta
A^\Lambda_{\nu}}=0\,\,\,\,&\Leftrightarrow& \qquad\qquad
\frac{1}{2}\,\epsilon^{\mu\nu\rho\sigma}\,D_\nu \,
G_{\rho\sigma}^{m}=\Omega^{mn}\,\frac{\delta{\cal
L}_{matter}}{\delta A^n_\mu}\equiv
\Omega^{mn}\,g\,j^{\mu}_n\,,\qquad\qquad\label{deltaa}
\end{eqnarray}
where $G^\Lambda\equiv H^\Lambda$. Equations (\ref{deltab}) imply
that on shell $H_\Lambda$ are dual to $H^\Lambda$. Since the
vector fields enter the scalar action only through the symplectic
invariant minimal couplings, the current $j^{\nu}_n$ will be
proportional to $\theta_n{}^\alpha$, and therefore, in virtue of
eq. (\ref{tad}), it will satisfy the property
\begin{eqnarray}
\theta_m{}^\alpha\,\Omega^{mn}\,j^{\nu}_n=0\,.\label{loc}\end{eqnarray}
Gauge invariance further requires that $D_\mu\,j^\mu_n=0$ on
shell.
 Let us discuss now the issue of locality. Although both electric and magnetic
vector fields take part to the action, the combinations of them
which are actually involved in the minimal couplings are well
defined since the corresponding magnetic currents vanish. Indeed
let $r={\rm rank}(\theta_n{}^\alpha)\le n_v$ and let us rewrite
the gauge connection in the following form, by using the fact that
 $X_n=\theta_n{}^\alpha\,t_\alpha$
\begin{eqnarray}
\Omega_\mu&=&A^\alpha_\mu\,t_\alpha\,\,\,;\,\,\,\,A^\alpha_\mu\equiv
A^n_\mu\,\theta_n{}^\alpha\,.
\end{eqnarray}
We see that only $A^\alpha_\mu$ are actually involved in the
minimal couplings and moreover that there are only $r$ independent
of them. If we contract both sides of eq. (\ref{deltaa}) by
$\theta_m{}^\alpha$ and use eq. (\ref{loc}) we see that the
$A^\alpha_\mu$ fields are well defined since the corresponding
field strengths
$G^\alpha_{\mu\nu}=\theta_n{}^\alpha\,G^n_{\mu\nu}$ satisfy the
Bianchi identities
\begin{eqnarray}
\epsilon^{\mu\nu\rho\sigma}\,D_\nu \,
G_{\rho\sigma}^{\alpha}&=&0\,.
\end{eqnarray}
The field equations derived from the variation of the tensor
fields $B_{\mu\nu\,\alpha}$ and the magnetic vector fields
$A_{\mu\,\Lambda}$ are non dynamic. This ensures the right number
of propagating degrees of freedom. For instance eqs.
(\ref{deltab}) can be solved to eliminate all the
$B_{\mu\nu\,\alpha}$ in favor of their scalar duals. The
propagating degrees of freedom
 can be ``distributed'' among the fields in different ways by
 performing different gauge fixings and then solving the
non-dynamic field equations. Until the gauge fixing is performed,
the theory is manifestly $G$--covariant.\par The embedding tensor
totally determines the form of the gauged Lagrangian. In
particular it determines the fermion shift matrices $S_{AB},\, N^{\mathcal{I} A}$
which enter the mass terms for gravitino and the fermion fields
and the scalar potential. In order to compute $S_{AB},\, N^{\mathcal{I} A}$ we
need first to introduce a scalar dependent $H$--tensor
$\mathcal{T}_{\hat{m}\hat{n}}{}^{\hat{p}}(\phi)$, called the
T--tensor \cite{dwn} (the hatted indices label the ${\bf 2\,n_v}$ as a
reducible representation of $H$). This tensor is obtained by
``dressing'' $X_{mn}{}^p$ by means of the vielbein
$\mathcal{V}_n{}^{\hat{n}}(\phi)$ of the coset manifold
$\mathcal{M}_{scal}$
\begin{eqnarray}
\mathcal{T}_{\hat{m}\hat{n}}{}^{\hat{p}}(\phi)&=&(\mathcal{V}^{-1})_{\hat{m}}{}^m\,
(\mathcal{V}^{-1})_{\hat{n}}{}^n\,\mathcal{V}_p{}^{\hat{p}}\,X_{mn}{}^p\,.
\end{eqnarray}
The shift matrices $S_{AB},\, N^{\mathcal{I} A}$ are then obtained by projecting
$\mathcal{T}$ into the relevant representations of $H$.
\section{M--theory compactified on a twisted seven--torus with fluxes}
Let us now discuss an application of the formalism discussed in
the previous section to the study of a specific compactification.
This will allow us to understand the relation between two dual
descriptions of the same theory. We shall consider M--theory
\cite{df,dft1,dft2} compactified on a twisted seven-torus in the
presence of fluxes. We start from the low-energy $D=11$
supergravity, which has $32$ supercharges and whose bosonic field
content consists of a graviton field
$\mathbb{V}_{\underline{\mu}}{}^{\underline{a}}$ and a 3--form
field $A^{(3)}_{\underline{\mu}\underline{\nu}\underline{\rho}}$,
where $\underline{\mu},\,\underline{\nu}=0,\dots,10$ and
$\underline{a},\underline{b}$ are the corresponding rigid indices.
The compactification on a twisted torus proceeds as follows. Let
us denote by $x^\mu$ ($\mu=0,\dots, 3$) the four dimensional
space--time coordinates, by $y^I$ ($I=4,\dots, 10$) the
coordinates on the internal seven--torus and by $a,b\dots$ the
rigid indices on the torus. The twisted seven--torus can be
locally described as a seven dimensional Lie group manifold
described by a basis of 1--forms $\sigma^I(y)=U(y)_J{}^I\,dy^J$
satisfying the following Cartan--Maurer equations
\begin{eqnarray}
d\sigma^I&=&\frac{1}{2}\,T_{JK}{}^I\,\sigma^J\wedge
\sigma^K\,,\label{twist}
\end{eqnarray}
where the structure constants of the group $T_{JK}{}^I$ define the
so called ``twist tensor''. This tensor is an instance of
\emph{geometric flux}. We shall restrict to ``volume preserving''
groups defined by the condition $T_{IJ}{}^J=0$. In \cite{alt} the
compactification on a twisted torus was alternatively described as
an ordinary toroidal compactification in the presence of an
internal \emph{torsion} $T_{JK}{}^I$.  \par The dimensional
reduction on this manifold is effected using for the various
fields the same ansatz as for the toroidal reduction except that
$dy^I$ are replaced by $\sigma^I$. In particular the ansatz for
the metric and the 3--form reads
\begin{eqnarray}
\mathbb{V}_{\underline{\mu}}{}^{\underline{a}}&=&\begin{cases}V_\mu{}^r\,dx^\mu\cr V^a=\phi^a_I \,(\sigma^I+A^I_\mu dx^\mu)\end{cases}\nonumber\\
        A^{(3)}&=&{\mathcal A}^{(3)}+B_I\wedge V^I+A_{IJ}\wedge V^I\wedge V^J+
C_{IJK}\wedge V^I\wedge V^J\wedge V^K\,.
\end{eqnarray}
where $A^I_\mu (x)$ are the seven Kaluza--Klein vectors and
$\phi^a_I(x)$ are the 28 moduli of the internal metric which span
the coset manifold ${\rm GL}(7,\mathbb{R})/{\rm
SO}(7)$.
The field content of the resulting four dimensional theory
consists in: The graviton field $V_\mu{}^r$, 28 vector fields
$A^I_\mu,\,A_{\mu\,IJ}$, 7 antisymmetric tensors $B_{\mu\nu\,I}$
and 63 scalar fields, 35 of which are the axions $C_{IJK}$
originating from the eleven dimensional 3--form, while the
remaining 28 are the $\phi^a_I$ fields. In the limit
$T_{JK}{}^I\rightarrow 0$ we are back to the ordinary toroidal
reduction which yields a version of the four dimensional ungauged
maximal supergravity featuring seven antisymmetric tensor fields.
As anticipated in section 2, in order for the global symmetry
group $G={\rm E}_{7(7)}$ of the ungauged theory to be manifest at
the level of field equations and Bianchi identities, the
antisymmetric tensors need to be dualized into scalar fields. The
global symmetry group of the ungauged Lagrangian is $G_e={\rm
GL}(7,\mathbb{R})$ and all fields and fluxes come in
representations of $G_e$, except for the metric moduli $\phi_I^a$
on which the action of $G_e$ is non-linear.  For instance the
twist tensor can be naturally associated with the representation
${\bf 140}_{+3}$ of $G_e$, where the subscript refers to the
grading with respect to the ${\rm O}(1,1)$ subgroup of $G_e$
acting as a rescaling of the internal volume. We can also switch
on form fluxes described by the v.e.v. of the eleven dimensional
4--form field strength along the internal directions ($g_{IJKL}$)
and along the four dimensional space--time directions
($\tilde{g}\,\epsilon_{\mu\nu\rho\sigma}$)
\begin{eqnarray}
F^{(4)}=dA^{(3)}+\tilde{g}\,\epsilon_{\mu\nu\rho\sigma}
\,dx^\mu\wedge dx^\nu\wedge dx^\rho\wedge dx^\sigma-g_{IJKL}\,
\sigma^I\wedge\sigma^J\wedge \sigma^K\wedge \sigma^L\,.
\end{eqnarray}
The background quantities $g_{IJKL}$ and $\tilde{g}$ are naturally
associated with the representations ${{\bf 35}^\prime}_{+5}$ and
${\bf 1}_{+7}$ of $G_e$ respectively. The $G_e$--representation of
the fields and fluxes is summarized in the Table below.
\begin{center}
\begin{tabular}{|c|c|c|c|c|c|c|c|c|}
  \hline
  Fields-fluxes &$V_\mu{}^a$ &$A^I_\mu$ & $A_{\mu\,IJ}$ & $B_{\mu\nu\,I}$ & $C_{IJK}$  & $T_{IJ}{}^K$ & $g_{IJKL}$ & $\tilde{g}$
  \\\hline
  ${\rm
GL}(7,\mathbb{R})$--reps. & ${\bf 1}_0$ & ${{\bf 7}}^\prime_{-3}$
& ${\bf 21}_{-1}$ & ${\bf 7}_{-4}$ &${\bf 35}_{+2}$ & ${\bf
140}_{+3}$ & ${{\bf 35}}^\prime_{+5}$  & ${\bf 1}_{+7}$  \\
\hline
\end{tabular}\label{tab1}
\end{center}
We can find the above representations in the branching of the
relevant ${\rm E}_{7(7)}$ representations with respect to $G_e$
\begin{eqnarray}
{\bf 56}&\rightarrow &  {{\bf 7}}^\prime_{-3}+{\bf 21}_{-1}+{\bf
7}_{+3}+ {{\bf 21}}^\prime_{+1}\,,\label{branch56}\\
{\bf 133}&\rightarrow &{\bf 48+1}_{0}+{\bf 35}_{+2}+ {{\bf
7}}^\prime_{+4}+ {\bf 35}^\prime_{-2}+{\bf 7}_{-4}\,,\label{branch133}\\
{\bf 912}&\rightarrow &{\bf 1}_{-7}+{\bf 1}_{+7}+{\bf 35}_{-5}+
{{\bf 35}}^\prime_{+5}+ ( {{\bf 140}}^\prime+ {{\bf
7}}^\prime)_{-3}+ ({\bf 140}+{\bf 7})_{+3}+{\bf 21}_{-1}+ {{\bf
21}}^\prime_{+1}+\nonumber\\&&{\bf 28}_{-1}+ {{\bf
28}}^\prime_{+1}+{\bf 224}_{-1}+ {{\bf
224}}^\prime_{+1}\label{branch912}\,.
\end{eqnarray}
In the branching (\ref{branch56}) the representations ${\bf
7}_{+3}$ and $ {{\bf 21}}^\prime_{+1}$ describe the vector fields
$\tilde{A}_{\mu\, I},\,\tilde{A}^{IJ}_\mu$ dual to $A^I_\mu$ and
$A_{\mu\,IJ}$ respectively. Recall that in the formulation
discussed in the previous section, the theory features all 70
scalar fields together with $133$ tensor fields, which include
their ``duals''. The branching (\ref{branch133}) can be used  to
identify the scalar fields transforming linearly under $G_e$,
which are $C_{IJK}$ and the scalars $\tilde{B}^I$ in the $ {{\bf
7}}^\prime_{+4}$, dual to $B_{\mu\nu\,I}$. It can also be used to
identify the tensor fields $B_{\mu\nu\,I}$ in the $ {{\bf
7}}_{-4}$, which, as we shall see,  are the only tensor fields
entering the Lagrangian.
 Finally the branching of the
${\bf 912}$ is useful in order to identify the fluxes which are
present in the compactification under consideration. Indeed the
most general gauging of maximal supergravity is described by an
embedding tensor transforming in the ${\bf 912}$ representation of
${\rm E}_{7(7)}$. In the right hand side of eq. (\ref{branch912})
we find, among the various representations, those defining the
fluxes which characterize the compactification we are considering,
thus confirming that fluxes enter the lower--dimensional gauged
supergravity as components of the embedding tensor which define
the gauge group.  Therefore in order to construct the gauge
algebra of the theory we just need to restrict the embedding
tensor to the representations ${\bf 140}_{+3}$, $ {{\bf
35}}^\prime_{+5}$ and ${\bf 1}_{+7}$. Group theory will do the
rest by completely determining the gauge algebra and then, through
the gauging procedure, the whole $\mathcal{L}_g$. The gauge
connection has the form
\begin{eqnarray}
\Omega_\mu &=&\tilde{A}^{MN}_{\mu}\, W_{MN}+A_{MN\mu}\,
W^{MN}+A^M_\mu\, Z_M\,,
\end{eqnarray}
where the gauge generators $W_{MN},\,W^{MN}$ and $Z_N$ close the
following algebra
\begin{eqnarray}
\left[X_n,\,X_m\right]=-X_{nm}{}^p\,X_p&\Leftrightarrow &
\begin{cases}\left[Z_M,\,Z_N\right]=T_{MN}{}^P\,Z_P+g_{MNPQ}\,W^{PQ}+\tilde{g}\,W_{MN}\,,\cr
 \left[Z_M,\,W^{PQ}\right]=2\,T_{MR}{}^{[P}\,W^{Q]R}
 +g_{MM_1M_2M_3}\,\epsilon^{M_1M_2M_3PQRS}\,W_{RS}\,,\cr
 \left[Z_M,\, {W_{PQ}}\right]=
 T_{PQ}{}^L\,W_{ML}\,,\cr
 \left[W^{IJ},\,W^{KL}\right]= {-3\,
T_{I_1I_2}{}^{[K}\,W_{I_3I_4}\epsilon^{L]IJI_1\dots
I_4}}\end{cases}\nonumber
\end{eqnarray}
all other commutators being zero. The locality condition
(\ref{tad}), which is also the condition for the above algebra to
close inside the algebra of ${\rm E}_{7(7)}$, amounts to requiring
that
\begin{eqnarray}
T_{[MN}{}^P\,T_{Q] P}{}^L=0\,,\\
T_{[MN}{}^P\,g_{QLR]P}=0\,.
\end{eqnarray}
The first condition is nothing but the Jacobi identity for the
algebra (\ref{twist}).\par We still have a redundancy of fields
and of gauge invariance. Let us now discuss two relevant gauge
fixings. The magnetic components of the embedding tensor are given
by the twist tensor: $\theta^{\Lambda\,\alpha}\equiv
\theta_{MN}{}^N=T_{MN}{}^N$. They contract only the tensor fields
$B_{\mu\nu\,I}$, out of the $B_{\mu\nu\,\alpha}$, which correspond
to the ${\rm E}_{7(7)}$ isometries $t_I$ acting as translations on
the Peccei-Quinn scalars $\tilde{B}^I$. Suppose that $T_{MN}{}^N$,
as a $21\times 7$ matrix, has maximal rank $7$. Note then that the
``magnetic'' vector fields $\tilde{A}^{MN}_\mu$ enter the
Lagrangian only in the seven independent combinations
$\tilde{A}^M_\mu\equiv T_{NP}{}^M\,\tilde{A}^{NP}_\mu$. We can
write the Stueckelberg term in the covariant derivative of
$\tilde{B}^M$ as follows
\begin{eqnarray}
D_\mu\tilde{B}^M&=&\partial_\mu
\tilde{B}^M+T_{NP}{}^M\,\tilde{A}^{NP}_\mu+\dots=\partial_\mu
\tilde{B}^M+\tilde{A}^M_\mu+\dots\,,
\end{eqnarray}
and eliminate $\tilde{B}^M$ by fixing the gauge freedom on $\tilde{A}^M_\mu$. After doing so, we can use one
of the non--dynamic field equations, which has the following form
\begin{eqnarray}
\epsilon^{\mu\nu\rho\sigma}\,   \partial_{\nu}B_{\rho\sigma I}
\propto g_{IJ}\,\tilde{A}^{J\mu}+\dots\,,
\end{eqnarray}
to eliminate $\tilde{A}^M_\mu$ in favor of $B_{\rho\sigma I}$. The
resulting gauge fixed theory is the one obtained in \cite{df} by direct
dimensional reduction and contains $7$ antisymmetric tensor fields
and $63$ scalar fields. This model features only the electric
vector fields $A_{\mu\,MN},\,A^M_\mu$. \par We can also perform a
different gauge fixing. Let us split the $21$ vector fields
$A_{\mu\,MN}$ into seven vector fields $A_{\mu\, M}$ and $14$
orthogonal components $A^\prime_{\mu\,MN}$
\begin{eqnarray}
A_{\mu\,MN}&=&T_{MN}{}^P\,A_{\mu\, P}+A^\prime_{\mu\,MN}\,.
\end{eqnarray}
The field strength $H_{\mu\nu\,MN}$ contains $dA_{ M}$ and $B_M$
in the following combination
\begin{eqnarray}
H_{\mu\nu\,MN}&=&T_{MN}{}^P\left(\partial_\mu A_{\nu\,
P}-\partial_\nu A_{\mu\, P}-B_{\mu\nu\,P}\right)+\dots\,.
\end{eqnarray}
We can therefore fix the tensor gauge transformation associated
with $B_{\mu\nu\,M}$ by ``eating'' the seven $A_{\mu\, P}$.  Then,
by using the non-dynamic equations (\ref{deltab}), which read
\begin{eqnarray}
F_{\mu\nu}{}^{MN}+\epsilon^{MNL_1\dots L_4 P}\,g_{L_1\dots
L_4}\,B_{\mu\nu\,P}&\propto& \epsilon_{\mu\nu\rho\sigma}\,
\frac{\delta{\cal L}}{\delta F_{\rho\sigma\,MN}}\,,
\end{eqnarray}
 the tensor fields $B_{\mu\nu\,P}$ can be eliminated in favor of
$\tilde{A}^{MN}_\mu $ (or equivalently of $\tilde{A}^M_{\mu}$).
The resulting gauge fixed theory, constructed in \cite{dft1,dft2}, features
$70$ scalar fields, no antisymmetric tensor fields, and $28$
electric vector fields consisting of the seven $A^M_\mu $, the 14
independent vectors out of the $A^\prime_{\mu\,MN}$ and the seven
$\tilde{A}^M_{\mu}$, which where originally described as magnetic.
This gauge fixing has therefore implied a symplectic rotation
\cite{s}. This is precisely the transformation $E$ in (\ref{E})
which is required to set the magnetic components of the embedding
tensor to zero. It is straightforward to generalize the above
discussion to the case in which the rank of $T_{MN}{}^N$, as a
$21\times 7$ matrix, is not maximal.
\par Let us finally comment on the vacua of the theory. These are
defined as the points in the scalar manifold which extremize the
scalar potential $V(\phi)$. The explicit expression of $V(\phi)$
was found in \cite{dft2,dp}. It consists of three terms
\begin{eqnarray}
V&=&V_E+V_K+V_{C-S}\,,\nonumber\\
V_E&=&\frac{1}{V_7}\,\left(2\, G^{KL}\, T_{KJ}{}^I\,
T_{LI}{}^J+G_{II'}\,G^{JJ'}\,
G^{KK'}\,T_{JK}{}^I\,T_{J'K'}{}^{I'}\right)\,,\nonumber\\
V_K&=&\frac{3}{16}\,\frac{1}{7!}\,\frac{1}{V_7}\,(g_{IJKL}+\frac{3}{2}\,T_{[IJ}^R\,C_{KL]R})(g_{MNPQ}+
\frac{3}{2}\,T_{[MN}^R\,C_{PQ]R})\, G^{IM}\, G^{JN}\, G^{KP}\,
G^{LQ}\,,\nonumber\\
V_{C-S}&=& \frac{1}{6}\,\frac{1}{V_7^3}\, \left(
C_{IJK}\,(g_{LPQR}+\frac{3}{4}\,T_{[LP}^N\,C_{QR]N})\,\epsilon^{IJKLPQR}+\tilde{g}\right)^2\,.\label{VV}
\end{eqnarray}
which come from the ten dimensional Einstein--Hilbert term, the
kinetic term of the 3--form and the Chern--Simons term
respectively ($V_7$ being the volume of the internal manifold).
Note that $V_K$ and $V_{C-S}$ are always positive definite while
$V_E$ is not. Extremizing $V$ with respect to $C_{IJK}$ we find
the following conditions
\begin{eqnarray}
g_{IJKL}+\frac{3}{2}\,T_{[IJ}^R\,C_{KL]R}&=&0\,,\label{cijk}
\end{eqnarray}
which admit a solution $C_{IJK}\equiv C_{IJK}^{(0)}$ in the axions
only for certain choices of $g_{IJKL}$.  It is straightforward to
show that this potential is always non--positive at its critical
points. Vacua with positive cosmological constant are thus ruled
out. In \cite{ft} it was shown that $V$ can at most vanish at its
critical points, thus ruling out also vacua with anti--de Sitter
geometry. It is interesting to consider the choices of fluxes
which allow for vacua with vanishing cosmological constant (called
``flat'' vacua). They define instances of the so called ``flat''
models which were extensively studied in the literature. It was
shown in \cite{dft2} that Minkowski vacua correspond to critical
points at which the three terms in $V$ vanish separately
\begin{eqnarray}
\mbox{Minkowski vacua}&\Rightarrow & V_E=V_K=V_{C-S}=0\,.
\end{eqnarray}
The vanishing of $V_{C-S}$, in particular, implies the following
condition on $\tilde{g}$
\begin{eqnarray}
\tilde{g}&=&\frac{3}{4}\,C^{(0)}_{IJK}\,T_{[LP}^N\,C^{(0)}_{QR]N}\,\epsilon^{IJKLPQR}\,.\label{cijk2}
\end{eqnarray}
The effect of the form--fluxes in these models is thus only to fix
the vacuum value of the axions, while the mass spectrum is
determined by $V_E$ and the mass parameters are encoded in
$T_{MN}{}^P$. \par To make a concrete example, let us consider the
case in which $I=4,i$  ($i=5,\dots, 10 $),  with
$T_{IJ}{}^K=T_{4i}{}^j$, zero otherwise, and $g_{IJKL}=g_{4ijk}$,
zero otherwise. In this case $T_{4i}{}^j=M_j{}^i$ is chosen to be
an antisymmetric matrix of rank 3 which can be set in the form:
\begin{eqnarray}
M_i{}^j&=&\left(\begin{matrix}m_1\,\epsilon &0&0\cr
0&m_2\,\epsilon &0\cr 0&0&m_3\,
\epsilon\end{matrix}\right)\,\,\,;\,\,\,\,\epsilon=\left(\begin{matrix}0&1\cr
-1&0\end{matrix}\right)\,.\label{skewt}
\end{eqnarray}
In this context  the equation (\ref{cijk}) fixes all $C_{ijk}$
fields but not the $C_{4ij}$ scalars. The $C_{4ij}$ fields give
masses to the $A_{\mu\,ij}$ vector fields with the exception of
the three entries $(ij)=(5,6),(7,8),(9,10)$. Therefore three of
the $C_{4ij}$ scalar remain massless moduli. The $G_{IJ}$--sector
gives, as discussed in reference \cite{ss}, four additional
massless scalars, two of which are the volume $V_7$ and $G_{44}$
and two other come from internal components of the metric.\par If
one further discusses the spectrum of the remaining fields, the
six vectors $A_{\mu\,4i}$ are eaten by the six antisymmetric
tensors $B_{\mu\nu\,i}$. An additional massless scalar comes from
the massless 2--form $B_{\mu\nu\,4}$ and finally an additional
massless vector come from the $A^4_\mu$ Kaluza--Klein vector. The
other six $A^i_\mu$ vectors become massive because of the twisting
of the torus. We conclude that in this theory there are always
eight massless scalars and four massless vectors, in agreement
with \cite{ss}.\par The fact that in all the models featuring
Minkowski vacua the form-fluxes never contribute to the physical
spectrum can be understood from a different perspective. The
fluxes $g_{IJKL}$ and $\tilde{g}$ for which eqs. (\ref{cijk}) and
(\ref{cijk2}) admit a solution $C^{(0)}_{IJK}$, can be seen as
generated by acting on $T_{MN}{}^P$ by means of an ${\rm
E}_{7(7)}$ duality transformation in the ${\bf 35}_{+2}$
representation on the right hand side of (\ref{branch133}),
parametrized by $C^{(0)}_{IJK}$. All these models therefore lie in
the same ${\rm E}_{7(7)}$ duality orbit as the model with
$g_{IJKL}=\tilde{g}=0$. If a suitable discrete form of ${\rm
E}_{7(7)}$ (the $U$--duality) were an exact symmetry of the
fundamental quantum theory of gravity, as conjectured in
\cite{ht}, then all the flat models arising from the class of flux
compactifications considered here should be interpreted as
different descriptions of the same microscopic degrees of
freedom.\section{The issue of ``non--geometric'' fluxes} The
background quantities $g_{IJKL},\,\tilde{g},\,T_{IJ}{}^K$ seem to
exhaust all possible form--  and geometric fluxes which can be
switched on in an M--theory compactification on a torus. Since all
the other components of the embedding tensor, defined by the
representations on the right hand side of (\ref{branch912}), lead
to a consistent four dimensional gauged theory, we may wonder if
they have any interpretation in terms of higher dimensional
fluxes. All the representations in the decomposition of the ${\bf
912}$ are connected to each other by the action of ${\rm
E}_{7(7)}$, which is conjectured to encode all known string
dualities. Therefore we can interpret the ${\rm
GL}(7,\,\mathbb{R})$ representations in the ${\bf 912}$, which do
not correspond to $g_{IJKL},\,\tilde{g},\,T_{IJ}{}^K$, as the
``duality image'' of the known form-- and geometric fluxes. Most
of these new background quantities can not be described in the
context of dimensional reduction on some compact manifold with
some global geometrical structure and therefore are called
``non--geometric'' fluxes. To give an example let us perform the
toroidal compactification of M--theory to four dimensions in two
steps: First compactify it on an $S^1$ along the eleventh
dimension $x^{10}$ and then compactify the resulting theory, which
is Type IIA superstring theory in ten dimensions, on a six torus
down to four dimensions
\begin{eqnarray}
\mbox{M--theory}&\stackrel{S^1(x^{10})}{\longrightarrow}&\mbox{
Type IIA superstring
($D=10$)}\stackrel{T^6}{\longrightarrow}\mbox{$D=4$
supergravity}\,.
\end{eqnarray}
The manifest symmetry of the resulting four dimensional Lagrangian
will be the subgroup ${\rm SL}(2,\mathbb{R})\times {\rm
GL}(6,\mathbb{R})$ of ${\rm E}_{7(7)}$, in particular the original
${\rm GL}(7,\mathbb{R})$ of the seven--torus is broken to ${\rm
O}(1,1)\times{\rm GL}(6,\mathbb{R})$, where ${\rm
GL}(6,\mathbb{R})$ acts on the metric moduli of the six--torus.
The geometric flux $T_{IJ}{}^K$ gives rise, upon reduction on
$S^1$, to the following quantities (for the sake of simplicity we
shall write, together with the ${\rm
SL}(6,\mathbb{R})$--representation, only the grading with respect
to the ${\rm O}(1,1)$ in ${\rm GL}(7,\mathbb{R})$) :
\begin{eqnarray}
T_{IJ}{}^K &\rightarrow & T_{uv}{}^w\,({\bf
84}_{+3}),\,T_{uv}{}^{10}\,({\bf 15}_{+3}),\,T_{10u}{}^{v}\,({\bf
35}_{+3}),\,(T_{uv}{}^{v}-T_{u 10}{}^{10})\,({\bf 6}_{+3})\,,
\end{eqnarray}
where $u,v=4,\dots, 9$,  $T_{uv}{}^w$ is the twist tensor defining
a ``twisted'' six--torus and $T_{uv}{}^{10}$ can be viewed as the
form--flux associated with the field strength of the R-R 1--form
field in Type IIA superstring theory. Each flux, being identified
with components  of the ${\bf 912}$ representation, is associated
with a  definite ${\rm E}_{7(7)}$ weight \cite{alt}, see appendix
A, so now we can study the effect of dualities on them.
$T$-duality is the perturbative equivalence between two string
theories: One compactified on a circle of radius $R$ and the other
compactified on a circle of radius $R^\prime=1/R$
($\alpha^\prime=1$). The most general $T$--duality transformation
in superstring theory compactified on a six--torus is described by
the discrete group ${\rm O}(6,6;\,\mathbb{Z})$, where the
restriction to the integer numbers is required by the boundary
conditions on the coordinates of the torus. In particular
$T$-duality transformations along an odd number of directions are
represented by elements of ${\rm O}(6,6;\,\mathbb{Z})$ with
negative determinant. Let us consider the effect of  T-duality
transformations on the geometric flux $T_{uv}{}^w$. If we apply on
$T_{uv}{}^w$ first a $T$--duality along $y^v$ ($T^{(v)}$) and then
one along $y^u$ ($T^{(u)}$)we obtain the following quantities
\begin{eqnarray}
T_{uv}{}^w\,\,({\bf 84}_{+3})&\stackrel{T^{(v)}}{\longrightarrow
}& Q_u{}^{vw}\,\,({{\bf
84}}^\prime_{+1})\stackrel{T^{(u)}}{\longrightarrow
}\,\,R^{uvw}({\bf 20}_{-1})\,.\label{htqr}
\end{eqnarray}
As observed in \cite{stw} the new quantities $Q_u{}^{vw}$ and
$R^{uvw}$ are instances of ``non--geoemtric'' fluxes. However the
corresponding representations ${{\bf 84}}^\prime_{+1}$ and ${\bf
20}_{-1}$ are contained in the ${\rm
GL}(7,\mathbb{R})$--representations  ${\bf 224}^\prime_{+1}$ and
${\bf 224}_{-1}$ in the branching (\ref{branch912}) of the ${\bf
912}$.  By restricting the embedding tensor to these
representations one can construct the whole four dimensional
supergravity originating from this generalized
flux-compactification. This is an example of how the
representations appearing the branching (\ref{branch912}) can be
related to each other by the action of string dualities.

\section{The $D=5\rightarrow D=4$ Scherk-Schwarz reduction and ``non--geometric'' fluxes}
 The Scherk--Schwarz (S-S) reduction on a circle from five to four dimensions was originally studied in \cite{ss,css}
 as a possible mechanism for producing an effective four dimensional supergravity featuring spontaneous supersymmetry
 breaking at various scales. It represents a generalized type of dimensional reduction
 in which the ansatz for the five dimensional fields, on a space-time of the form $\mathbb{R}^{1,3}\times
 S^1$,
 contains a dependence on the internal $S^1$ coordinate $y$ through a
 global symmetry transformation of the five dimensional
 Lagrangian, called Scherk--Schwarz ``twist''
 \begin{eqnarray}
\Phi(x^\mu,y)&=&e^{M\,y}\cdot \phi(x^\mu)\,,
\end{eqnarray}
where $e^{M\,y}$ is the twist matrix and $M$ is a global symmetry
generator of the five dimensional theory which has a non-trivial
action on the field $\Phi$\footnote{In the literature distinction
is made between reductions in which the twist is taken in the
symmetry group ${\rm SL}(n,\mathbb{R})$ of the five dimensnional
theory, once the latter is interpreted as originating from a
dimensional reduction on an n--torus \cite{ss}, and reductions in
which the twist is a generic global symmetry of the theory itself
\cite{css}. In the former case the reduction is referred to as
``Scherk--Schwarz reduction'', while in the latter case as
``generalized Scherk--Schwarz reduction'' or ``duality twist''.
Here we shall not use this distinction.} . This property of $M$
guarantees that the dependence on $y$ ultimately disappears in the
four dimensional theory. However since $y$ has the dimension of an
inverse mass, $M$ has the dimension of a mass and will induce mass
deformations in the lower dimensional theory. They originate from
terms, in the $D=5$ Lagrangian, containing derivatives with
respect to $y$: $\partial^2_y\Phi(x,y)=M^2\cdot \Phi(x,y)$. If we
start from five dimensional \emph{maximal} (ungauged)
supergravity, whose Lagrangian has an ${\rm E}_{6(6)}$ global
symmetry group, we can perform a S-S reduction by taking as $M$
any generator of ${\rm E}_{6(6)}$. The resulting four dimensional
supergravity is a gauged supergravity, as was first shown in
\cite{adfl}. This model is an instance of a ``no--scale''
supergravity as it features a
 non--negative scalar potential. The only possible vacua are of Minkowski type
 and are defined by the points in the scalar manifold in which the potential vanishes. These points exist
only if $M^T=-M$, namely if $M$ is a generator of the maximal
compact subgroup ${\rm USp}(8)$ of ${\rm E}_{6(6)}$, in which case
the gauge group is called ``flat'' group. The embedding tensor for
this theory was constructed in \cite{dwst1}. It is defined by the
 ${\bf 78}_{+3}$ representation in the branching of the ${\bf 912}$ with respect to the
 subgroup ${\rm E}_{6(6)}\times {\rm O}(1,1)$ of ${\rm E}_{7(7)}$.
  In the basis of the ${\bf 56}$ in which the 28 electric
vector fields are $A^\Lambda_\mu=\{A^0_\mu,\,A^\lambda_\mu\}$, where
$A^\lambda_\mu$, $\lambda=1,\dots, 27$, are the dimensionally reduced five--
dimensional vectors in the ${\bf 27}_{-1}$ of ${\rm E}
_{6(6)}\times {\rm O}(1,1) $ and $A^0_\mu$ is the Kaluza--Klein
vector in the ${\bf 1}_{-3}$ of the same group, the embedding
tensor has just electric components $\theta_\Lambda^\sigma$ and
the gauge generators $X_\Lambda$ read:
\begin{eqnarray}
X_\Lambda &=&\begin{cases}X_0=\theta_{0,\lambda}{}^\delta\,t_{\delta}{}^\lambda\cr
X_\lambda=\theta_\lambda{}^\delta\,t_\delta\end{cases}\,\,\,\,;\,\,\,\,
\theta_{0,\lambda}{}^\delta =\theta_\lambda{}^\delta=M_\lambda{}^\delta\in {\rm E} _{6(6)}\,.
\end{eqnarray}
where $M_\lambda{}^\delta$ is the twist matrix depending in general on $78$
parameters, $t_\lambda{}^\delta$ are the ${\rm E} _{6(6)}$ generators, and
$t_\lambda$ are ${\rm E} _{7(7)}$ generators in the ${{\bf
27}}^\prime_{+2}$, according to the following branching of the
${\rm E} _{7(7)}$ generators with respect to ${\rm E}
_{6(6)}\times {\rm O}(1,1) $:
\begin{eqnarray}
{\bf 133}&\rightarrow & {\bf 78}_0+{\bf 1}_0+{{\bf
27}}^\prime_{+2} +{\bf 27}_{-2}\,.
\end{eqnarray}
In this case the relevant components of the gauge generators
$X_{\Lambda\,m}{}^n$ are:
\begin{eqnarray}
X_{0\lambda}{}^\delta &=&-X_{\lambda0}{}^\delta=-X_{0}{}^\delta{}_\lambda=X_\lambda{}^\delta{}_0=-M_\lambda{}^\delta\,\,\,\,\,;\,\,\,\,\,\,
X_{\lambda \delta \gamma}=M_\lambda{}^{\lambda^\prime}\,d_{\lambda^\prime \delta \gamma}\,,\label{Xss}
\end{eqnarray}
where $d_{\lambda \delta \gamma}$ denotes the three times symmetric invariant tensor
of the ${\bf 27}$ of ${\rm E} _{6(6)}$. To obtain eqs. (\ref{Xss})
we have used the properties
$(t_\lambda{}^\delta)^{\gamma_1}{}_{\gamma_2}=-(t_\lambda{}^\delta)_{\gamma_2}{}^{\gamma_1}=\delta^{\gamma_1}_\lambda\delta^\delta_{\gamma_2}-(1/27)\,\delta^{\gamma_1}_{\gamma_2}\delta^\delta_\lambda$,
$(t_\lambda)^\delta{}_0=-(t_\lambda)_0{}^\delta{}=\delta^\delta_\lambda$ and $(t_\lambda)_{ \delta \gamma}=d_{\lambda \delta \gamma}$.
The gauge algebra has the following structure:
\begin{eqnarray}
[X_0,\,X_\lambda]&=& M_\lambda{}^\delta\,X_\delta\,,
\end{eqnarray}
all other commutators vanishing. \par If $M $ is non--compact the
corresponding theory depends effectively only on six parameters
and the potential is of run--away type, namely there is no vacuum
solution. If, on the other hand, $M$ is compact, the theory has
Minkowski vacua and depends effectively on four mass parameters
$m_1,\,m_2,\,m_3,\,m_4$, since $M$ can always be reduced to an
element of the maximal torus of  ${\rm USp}(8)$. These mass
parameters fix the scale of spontaneous supersymmetry breaking,
which can yield an $N=6,4,2$ or $N=0$ effective theory.\par Can we
interpret this four dimensional spontaneously broken supergravity
as originating from an eleven dimensional flux compactification?
To this end let us describe the five dimensional theory as
originating from the compactification of M--theory on a
six--torus. This is done by branching the relevant ${\rm E}
_{6(6)}$--representations with respect to the ${\rm
SL}(2,\mathbb{R})\times {\rm SL}(6,\mathbb{R})$ subgroup of ${\rm
E} _{6(6)}$, ${\rm SL}(6,\mathbb{R})$ being, as usual, the group
acting on the metric moduli of the six--torus. In particular we
are interested in interpreting the embedding tensor of the model
in terms of higher dimensional fluxes. This is done by decomposing
the representation ${\bf 78}_{+3}$ of $\theta$ with respect to $
{\rm SL}(6,\mathbb{R})\times{\rm SL}(2,\mathbb{R})$
\begin{eqnarray}
\begin{CD}{\bf 78} @>>{{\rm SL}(6,\mathbb{R})\times {\rm SL}(2,\mathbb{R})}> ({\bf 35},{\bf 1}) + ({\bf 1},{\bf 3}) + ({\bf
20},{\bf 2})\end{CD} \label{sl6decomp}
\end{eqnarray}
From the detailed analysis in \cite{alt}, it follows that the
${\bf (35,1)}$ can be interpreted as the $T_{4i}{}^j$ components
of the twist tensor $T_{IJ}{}^K$. Indeed $T_{4i}{}^j$ is in
general a generator of $ {\rm SL}(6,\mathbb{R})$,  and flat vacua
occur only if this matrix is antisymmetric, namely if it is an $
{\rm SO}(6)$ generator. In this case it can always be brought to
the skew--diagonal form (\ref{skewt}) and will contribute three
mass parameters to the theory.  How about the remaining fourth
mass parameter? It will come from a compact ${\rm E} _{6(6)}$
twist which is not in an $ {\rm SL}(6,\mathbb{R})$-generator. Let
us have a closer look at the branching (\ref{sl6decomp}). The
representation ${\bf 20}$ of $ {\rm SL}(6,\mathbb{R})$ appears in
a doublet with respect to $ {\rm SL}(2,\mathbb{R})$. One component
of the doublet, which we shall call ${\bf 20}_+$, as it was shown
in \cite{alt}, can be interpreted as the components $g_{4ijk}$ of
$g_{IJKL}$. It corresponds to a nilpotent generator of ${\rm E}
_{6(6)}$ and thus, alone, it cannot contribute any mass parameter.
In fact, as we have seen in section 5, its only effect is to fix
the axions to certain values. Similarly the highest grading
component of the triplet ${\bf (1,3)}$, which we shall denote by
${\bf 1}_+$, can be identified with $\tilde{g}$. It also singles
out a nilpotent twist matrix $M$, this time corresponding to the
positive root $\alpha$ of $ {\rm SL}(2,\mathbb{R})$. In order to
get a compact matrix $M$, namely antisymmetric since we are always
working with real representations, it should be the combination of
an upper and a lower triangular matrix, namely of two shift
matrices with opposite gradings, corresponding to positive ${\rm
E} _{6(6)}$ roots and their negative respectively. Therefore the
fourth parameter can only arise from a combination of the ${\bf
20}_+$ with ${\bf 20}_-$ components of the ${\bf (20,2)}$, or from
a combination of the ${\bf 1}_+$ and ${\bf 1}_-$ components of the
${\bf (1,3)}$. The ${\bf 20}_-$ and the ${\bf 1}_-$ components do
not have an interpretation in terms of form--or geometric fluxes,
nevertheless they are obtained by acting on the known ${\bf 20}_+$
and ${\bf 1}_+$ respectively by means of the Weyl transformation
\cite{pope} associated with the root $\alpha$ of $ {\rm
SL}(2,\mathbb{R})$. This is a proper $U$--duality transformation
which mixes the internal radii of the torus (in the ten
dimensional string frame) $R_i$, together with the ten dimensional
dilaton field $\phi$. Its action can be deduced using the analysis
of \cite{alt}
\begin{eqnarray}
R_4^\prime &=&
e^{-\frac{\phi}{2}}\,V_6^{\frac{1}{2}}\,R_4^{\frac{1}{2}}\,,\nonumber\\
R_i^\prime
&=&e^{\frac{\phi}{2}}\,V_6^{-\frac{1}{2}}\,R_4^{\frac{1}{2}}\,R_i\,\,\,\,;\,\,\,\,i=5,\dots,9\,,\nonumber\\
e^{\phi^\prime}&=&e^{\frac{3}{2}\,\phi}\,V_6^{-\frac{1}{2}}\,R_4^{\frac{1}{2}}\,,\label{uduatra}
\end{eqnarray}
where $V_6=R_4\,R_5\,\dots R_9$ is the volume of the six torus,
see appendix A. To be able to interpret the fourth S-S parameter
therefore we need to include, apart from the form--fluxes, also
their U-duality image, in a single picture! This unifying picture
of compactification, accommodating at the same time U--dual
background quantities, could be provided by the so called
``U--folds''\cite{dh}, namely non--geometric manifolds having
U--duality transformations as transition functions. A similar
analysis, from a different perspective, was carried out in
\cite{dh}.\par An other interesting duality between different flux
compactifications was found in \cite{gargiulo} using the embedding
tensor description of gauged supergravity. In this paper it was
observed that if we restricted the embedding tensor to the
representation ${\bf (20,2)}$ in (\ref{sl6decomp}), upon an $N=4$
truncation of the $N=8$ supergravity, the resulting theory
coincides with the $N=4$ gauged supergravity describing Type IIB
superstring compactified on a $T^6/\mathbb{Z}_2$--orientifold in
the presence of R-R and NS-NS 3--form fluxes $F^{(3)},\,H^{(3)}$.
Indeed these fluxes transform in the ${\bf (20,2)}$ representation
of ${\rm SL}(6,\mathbb{R})\times {\rm SL}(2,\mathbb{R})$, though
in the Type IIB setting the ${\rm SL}(2,\mathbb{R})$ symmetry
group is interpreted as the global symmetry of the ten dimensional
theory. This is an instance of a duality between two seemingly
different theories:
 Type IIB superstring reduced on a
$T^6/\mathbb{Z}_2$--orientifold in the presence of 3--form fluxes
and a (truncation of a)  Scherk--Schwarz reduction from five dimensions, which has a more natural
description as originating from M--theory or Type IIA superstring.

\section{Conclusions and outlook}
In the present paper have reviewed a new description of gauged
supergravity in which the whole global (non--perturbative)
symmetry group $G$ of the ungauged Lagrangian is preserved at the
level of field equations and Bianchi identities. This description
includes the scalar fields together with their dual tensor fields,
the electric vector fields and their magnetic duals. All the
information about the gauge group is encoded in the a
$G$--covariant embedding tensor. We have also discussed some
applications of this description. The relevance of this
formulation is apparent if we consider flux compactifications,
since the internal fluxes naturally enter the lower dimensional
effective supergravity as components of the embedding tensor.
Form-- and  geometric fluxes, however, fit a restricted number of
components. All the remaining components, which are consistent
with the supersymmetry constraints, can be obtained from the known
fluxes by means of U--duality transformations, which include the
$T$ and $S$-dualities. Lifting the corresponding gauged models to
ten or eleven dimensions is still an open problem. Nevertheless
the embedding tensor formulation of gauged supergravity represents
an ideal laboratory in which to study the web of dualities
connecting the known flux compactifications with generalizations
thereof. For example the duality covariant formulation reviewed in
section 4 can be applied to the construction of the complete
mirror-symmetry-covariant four dimensional supergravity
description of Type II superstring compactified on ${\rm
SU}(3)\times {\rm SU}(3)$--structure manifold in the presence of
general fluxes \cite{glw2,micu}.  In this case the (electric)
embedding tensor, defined in \cite{heis}, which gauges (an abelian
subalgebra of) the Heisenberg isometry algebra of the $N=2$
quaternionic manifold, and which reproduces the form- and
geometric fluxes, has to be extended to include magnetic
components as well, in order to account for the non-geometric
fluxes. This has been done in \cite{heis2}, were the low-energy
gauged $N=2$ supergravity reproducing the general flux
compactification described in \cite{glw2} was constructed. It is
still an open question whether such compactifications can be
further generalized to yield a non-abelian gauging in four
dimensions.

\section{Acknowledgements}
M.T. would like to thank L. Andrianopoli, R. D'Auria, B. de Wit,
S. Ferrara, M. Gra\~na, J. Louis and H. Samtleben for useful
discussions.
 Work supported in part by the European
Community's Human Potential Program under contract
MRTN-CT-2004-005104 `Constituents, fundamental forces and
symmetries of the universe'.
\appendix
\section{Fluxes and ${\rm E}_{7(7)}$ weights}
In this appendix we illustrate is some detail how to associate the
known fluxes, in ten dimensional Type II superstring compactified
on a six-torus, with components of the embedding tensor which
defines the corresponding four dimensional gauged supergravity,
namely with elements of the ${\bf 912}$ representation of ${\rm
E}_{7(7)}$. An element of a Lie group $G$  representation  is
characterized by its transformation property under the action of
the maximal torus of $G$, generated by its Cartan subalgebra
(CSA). This property is encoded in the \emph{weights} of the
representation. Let $G$ be the global symmetry group of an
extended supergravity theory. Its maximal torus has a diagonal
action on the electric field strengths and their duals, and
therefore it is a symmetry of the ungauged Lagrangian, namely it
is contained in $G_e$. If fluxes are to be assigned, by
identification with components of the embedding tensor, to
$G$-representations, so as to restore on shell global $G$
invariance, they should couple in the action to the dilatonic
fields parametrizing the CSA of $G$ according to their weights. In
the maximal theory the $CSA$ of ${\rm E}_{7(7)}$ is seven
dimensional and is parametrized, from the Type II superstring
point of view, by the six radial moduli of the internal torus
$R_i=e^{\sigma_i}$ ($i=4,\dots,9$) and by the ten dimensional
dilaton $\phi$. The bosonic zero-modes of Type II superstring
theory consist in the ten dimensional dilaton $\phi$, the metric
$\mathbb{V}_{\hat{\mu}}{}^{\hat{a}}$
($\hat{\mu},\,\hat{\nu}=0,\dots,9$ and
$\hat{a},\,\hat{b}=0,\dots,9$ are the curved and rigid ten
dimensional indices respectively), a NS-NS 2--form $B_{(2)}$, odd
R--R forms, $C_{(1)},\,C_{(3)}$, in Type IIA theory and even R--R
forms, $C_{(0)},\,C_{(2)},\,C_{(4)}$, in Type IIB. The ansatz for
the metric in the string frame reads
\begin{eqnarray}
\mathbb{V}_\mu{}^r=
e^{\phi_4}\,V_\mu{}^r\,\,\,;\,\,\,\,\mathbb{V}^{\hat{u}}=\phi_u{}^{\hat{u}}\,(dy^u+A^u_\mu\,dx^\mu)\,,
\end{eqnarray}
where $u,v=4,\dots, 9$ and $\hat{u},\hat{v}=4,\dots, 9$ are the
curved and rigid indices on the six torus respectively,
$V_\mu{}^r$ is the four dimensional metric in the four dimensional
Einstein frame, $A^u_\mu$ are the six Kaluza Klein vectors,
$\phi_4=\phi-\frac{1}{2}\,\sum_u\sigma_u$ is the four dimensional
dilaton and $\phi_u{}^{\hat{v}}$ are the metric moduli of the
internal torus, which can be identified with the coset
representative of ${\rm GL}(6,\mathbb{R})/{\rm SO}(6)$. By
suitably fixing the ${\rm SO}(6)$ symmetry we can adopt the
solvable Lie algebra representation of the manifold ${\rm
GL}(6,\mathbb{R})/{\rm SO}(6)$ \cite{cj,sla,dd} and write
$\phi_u{}^{\hat{v}}$ in the form
\begin{eqnarray}
\phi_u{}^{\hat{v}}&\equiv &
U\,e^{\sum_{u=4}^9\sigma_u\,H_{\epsilon_u}}\,,\nonumber\\
U&=&\prod_{u<v}e^{\gamma_u{}^v\,E_u{}^v}\,\,\,\,\,(\mbox{no
summation})\,,\label{Uu}
\end{eqnarray}
where $\epsilon_u$ is an orthonormal basis of vectors, $E_u{}^v$
are the ${\rm SL}(6,\mathbb{R})$ shift generators corresponding to
the positive root $\epsilon_u-\epsilon_v$ and $\gamma_u{}^v$ are
the moduli parametrizing the off diagonal components of the
internal metric. The internal metric will read
$G_{uv}=-\sum_{\hat{w}}\phi_u{}^{\hat{w}}\phi_v{}^{\hat{w}}$. Let
us define a representative of the maximal torus of ${\rm
E}_{7(7)}$ to have the form $\exp(H_{\vec{h}})$, where $\vec{h}$
is defined as
\begin{eqnarray}
\vec{h}(\sigma,\phi)&=&\sum_{u=4}^9\sigma_u\,\epsilon_u-\sqrt{2}\,\phi_4\,\epsilon_{10}=
\sum_{u=4}^9\hat{\sigma}_u\,(\epsilon_u+\frac{1}{\sqrt{2}}\,\epsilon_{10})-\frac{1}{2}\,\,\phi\,a\,,\nonumber\\
a&=&-\frac{1}{2}\,\sum_{u=4}^9\sigma_u\,\epsilon_u+\frac{1}{\sqrt{2}}\,\epsilon_{10}\,,\label{hvect}
\end{eqnarray}
where $\epsilon_I=(\epsilon_u,\epsilon_{10})$ is an orthonormal
basis of seven dimensional vectors and
$\hat{\sigma}_u=\sigma_u-\phi/4$ are the radial moduli in the ten
dimensional Einstein frame. The four dimensional Lagrangian,
resulting from the dualization of the 2-forms to scalar fields,
will contain the following terms
\begin{eqnarray}
e^{-1}\mathcal{L}_{scal}&=&\frac{1}{2}\partial_\mu\vec{h}\cdot\partial^\mu\vec{h}+\frac{1}{4}\,\sum_{u,v}\,e^{-2\,\alpha^{B}_{uv}\cdot
\vec{h}}\,\partial_\mu B_{uv}\,\partial^\mu
B_{uv}+\frac{1}{2}\,\sum_{u<v}\,e^{-2\,\alpha_u{}^v\cdot
\vec{h}}\,\partial_\mu\gamma_u{}^v\partial^\mu\gamma_u{}^v+\nonumber\\
&&+\sum_k\frac{1}{2k!}\,\sum_{u_1,\dots,u_k}\,e^{-2\,\alpha^C_{u_1\dots
u_k}\cdot \vec{h}}\,\partial_\mu C_{u_1\dots u_k}\partial^\mu
C_{u_1\dots u_k}+\frac{1}{2}\,e^{-2\,\alpha^{B}\cdot
\vec{h}}\,\partial_\mu B\,\partial^\mu B\dots\,,\label{lscal}\nonumber\\&&\\
e^{-1}\mathcal{L}_{vec}&=&-\sum_{u}e^{-2\,W^u\cdot \vec{h}
}\,\partial_{[\mu}A^u_{\nu]}\partial^{[\mu}A^{\nu]u}-\sum_{u}e^{-2\,W_u^B\cdot
\vec{h}
}\,\partial_{[\mu}B_{\nu]u}\partial^{[\mu}B^{\nu]}{}_u-\nonumber\\&&-
\sum_k\frac{1}{(k-1)!}\,\sum_{u_1,\dots,u_{k-1}}\,e^{-2\,W^C_{u_1\dots
u_{k-1}}\cdot \vec{h}}\,\partial_{[\mu}C_{\nu]u_1\dots
u_{k-1}}\partial^{[\mu}C^{\nu]}{}_{u_1\dots
u_{k-1}}+\dots\,.\nonumber\\&&\label{lvec}
\end{eqnarray}
where the internal indices of the scalar and vector fields are
``dressed'' with the matrix $U$ in eq. (\ref{Uu}) and the ellipses
comprise the non linear couplings deriving from the Chern-Simons
terms in the definition of the ten dimensional field strengths.
The 2-form $B_{\mu\nu}$ has been dualized to the axion $B$ while
in the Type IIA theory the tensors $C_{\mu\nu u}$ were dualized to
$\epsilon_{u u_1\dots u_5 }\,C_{u_1\dots u_5}$ and in the Type IIB
theory $C_{\mu\nu}$ was dualized to the scalar $C_{4\dots 9}$. The
range of values of $k$ in the summations in (\ref{lscal}) and
(\ref{lvec}) is: $k=1,3,5$ in Type IIA and $k=0,2,4,6$ in Type
IIB. The seven dimensional vectors $\alpha$ and $W$ in the
exponential factors of (\ref{lscal}) and (\ref{lvec}) have the
form
\begin{eqnarray}
\alpha_u{}^v&=&\epsilon_u-\epsilon_v\,\,;\,\,\,\alpha^B_{uv}=\epsilon_u+\epsilon_v\,\,;\,\,\,\alpha^B=\sqrt{2}\,\epsilon_{10}\,\,;\,\,\,
\alpha^C_{u_1\dots u_k}=a+\epsilon_{u_1}+\dots
+\epsilon_{u_k}\,,\label{alpha}\\
W^u&=&-\epsilon_u-\frac{1}{\sqrt{2}}\,\epsilon_{10}\,\,;\,\,\,W^B_u=\epsilon_u-\frac{1}{\sqrt{2}}\,\epsilon_{10}\,\,;\,\,\,W^C_{u_1\dots
u_{k-1}}=a+\epsilon_{u_1}+\dots
+\epsilon_{u_{k-1}}-\frac{1}{\sqrt{2}}\,\epsilon_{10}\,,\nonumber\\&&\label{ww}
\end{eqnarray}
If we define the simple roots of the $\mathfrak{e}_{7(7)}$ algebra
to be of the form
\begin{eqnarray}
\alpha_{u-3}&=&\epsilon_u-\epsilon_{u+1}\,\,\,(u=4,\dots,
8)\,;\,\,\,\,\alpha_6=\epsilon_8+\epsilon_9\,;\,\,\,\alpha_7=\begin{cases}a&
\mbox{Type IIB}\cr a+\epsilon_9 & \mbox{Type IIA}
\end{cases}\,,\label{iiaiib}
\end{eqnarray}
 the vectors in (\ref{alpha}) are the $\mathfrak{e}_{7(7)}$ positive
roots while those in (\ref{ww}), together with their opposite $-W$
(corresponding to the magnetic vector fields), are the weights of
the ${\bf 56}$ representation (in the Type IIB description, the
weights $W^C_{u_1u_2u_3}$ are $20$ and correspond to the vectors
$C_{\mu u_1u_2u_3}$ originating from the 4-form; in virtue of the
property of the 5-form field strength of being self dual, these 20
weights already include 10 weights corresponding to electric
vector fields and their opposite associated with the magnetic
duals). We may follow a similar strategy in order to associate
fluxes with $\mathfrak{e}_{7(7)}$ weights, namely read off the
weight from the dilaton dependence of the term in the action of
the form $(\mbox{flux})^2$. The relevant terms in the Lagrangian
are
\begin{eqnarray}
e^{-1}\,\mathcal{L}&=& \sum_{u,v,w}\left(e^{-2W^T_{uv}{}^w\cdot
\vec{h}}\,(T_{uv}{}^w)^2-\frac{1}{12}\,e^{-2W^H_{uvw}\cdot
\vec{h}}\,(H^{(3)}_{uvw})^2\right)+\nonumber\\&&+\sum_k(-1)^{(k+1)}\frac{1}{2\,(k+1)!}\,\sum_{u_1\dots
u_{k+1}}e^{-2\,W^F_{u_1\dots u_{k+1}}\cdot
\vec{h}}\,(F^{(k+1)}_{u_1\dots u_{k+1}})^2+\dots\,,\label{lflux}
\end{eqnarray}
where the term containing $(T_{uv}{}^w)^2$ is part of the
Scherk-Schwarz potential \cite{ss} which, in the case of M-theory
compactification on a twisted torus, is described by $V_E$ in
(\ref{VV}). The rank $k$ in the summation over the R-R fluxes have
the values $k=-1,1,3,5$ in Type IIA theory, corresponding to the
internal components of the forms
$F^{(0)},\,F^{(2)},\,F^{(4)},\,F^{(6)}$, and $k=0,2,4$ in Type IIB
theory, corresponding to the field strengths
$F^{(1)},\,F^{(3)},\,F^{(5)}$. We may also consider R-R fluxes
with four space-time indices which do not explicitly break Lorentz
invariance. By performing the dimensional reduction, we find the
general field-strength--weight correspondence:
\begin{eqnarray}
H^{(3)}_{u_1u_2u_3}&\leftrightarrow &
W^H_{u_1u_2u_3}=\epsilon_{u_1}+\epsilon_{u_2}+\epsilon_{u_3}+\frac{1}{\sqrt{2}}\,\epsilon_{10}\,,\nonumber\\
T_{u_1u_2}{}^{u_3}&\leftrightarrow &
W^T_{u_1u_2}{}^{u_3}=\epsilon_{u_1}+\epsilon_{u_2}-\epsilon_{u_3}+\frac{1}{\sqrt{2}}\,\epsilon_{10}\,,\nonumber\\
F^{(k+1)}_{\mu_1\dots\mu_\ell u_1\dots u_{s}}&\leftrightarrow &
W^F_{\mu_1\dots\mu_\ell u_1\dots
u_s}=-\frac{1}{2}\,\sum_u\epsilon_u+\epsilon_{u_1}+\dots+\epsilon_{u_s}+
\frac{2-\ell}{\sqrt{2}}\,\epsilon_{10}\,\,\,\,\,(\ell+s=k+1)\,.\nonumber\\&&\label{weightflux}
\end{eqnarray}
In the M-theory reduction on a torus the ${\rm O}(1,1)$ factor in
$G_e={\rm GL}(7,\mathbb{R})$ is generated by the Cartan operator
$H_\lambda$ where
\begin{eqnarray}
\lambda&=&\sum_u\epsilon_u+2\sqrt{2}\,\epsilon_{10}\,,
\end{eqnarray}
and the ${\rm O}(1,1)$ weight associated with the field strengths
in (\ref{weightflux}) are simply computed as the scalar product of
$\lambda$ with the corresponding weight $W$: $\lambda\cdot W$.
From the embedding of the ${\rm SL}(6,\mathbb{R})$ group,
corresponding to the six torus in the compactification of Type II
theories, inside ${\rm E}_{7(7)}$, we may deduce the ${\rm
SL}(6,\mathbb{R})$-representation  of each of the weights in
(\ref{weightflux}) and identify it, together with the relevant
${\rm O}(1,1)$ gradings, with representations in the branching of
the embedding tensor representation ${\bf 912}$. The embedding of
${\rm SL}(6,\mathbb{R})$ inside ${\rm E}_{7(7)}$ is defined by
identifying its simple roots with $\alpha_1\dots \alpha_5$.\par
Let us now consider the effect of dualities. An important role is
played by those dualities which are effected as Weyl
transformations $\sigma_\Delta$ with respect to a given weight
$\Delta$, whose action on a weight $W$ is defines as follows
\begin{eqnarray}
W&\longrightarrow &\sigma_\Delta (W)=W-2\,\left(\frac{W\cdot
\Delta}{\Delta\cdot \Delta}\right)\,\Delta\,.
\end{eqnarray}
If $\Delta$ is a root of $\mathfrak{e}_{7(7)}$ then this
transformation is an ${\rm E}_{7(7)}$ transformation and therefore
a symmetry of the theory. This is not always the case for the
known string dualities. Let us see what the effect of these
transformations is on the dilatonic scalars. Since $\sigma_\Delta$
is an orthogonal transformation in the Euclidean vector space of
the CSA of $\mathfrak{e}_{7(7)}$, we can rewrite the generic
exponent in (\ref{lscal}), (\ref{lvec}) and (\ref{lflux}) as
follows
\begin{eqnarray}
e^{-2\,W\cdot \vec{h}}&=&e^{-2\,\sigma_\Delta (W)\cdot
\sigma_\Delta(\vec{h})}\,,
\end{eqnarray}
namely the field which is described by the weight $W$ (which, for
a scalar field, is a positive root) in the original theory,
corresponds to the new weight $\sigma_\Delta (W)$ in the dual
theory, which features a new set of dilatonic scalars
$\sigma_u^\prime,\,\phi^\prime$, entering the dilatonic vector
$\sigma_\Delta(\vec{h})$. The relation between
$\sigma_u^\prime,\,\phi^\prime$ and $\sigma_u,\,\phi$ can be
deduced by the following condition
\begin{eqnarray}
\vec{h}(\sigma^\prime,\phi^\prime)&\equiv
&\sigma_\Delta(\vec{h}(\sigma,\phi))\,\,\,\Rightarrow\,\,\,\,\sigma^\prime=\sigma^\prime(\sigma,\phi)\,,\,\,\,\phi^\prime=\phi^\prime(\phi)\,.\label{htrans}
\end{eqnarray}
Let us consider as an example the $T$--duality $T^{(u)}$ along the
internal direction $y^u$. It is implemented by the Weyl
transformation $\sigma_{\epsilon_u}$ \cite{RRbh}, corresponding to
the vector $\Delta=\epsilon_u$. Since this vector is not an
$\mathfrak{e}_{7(7)}$ root, in general $T$--duality along an odd
number of direction is not an ${\rm E}_{7(7)}$ transformation. If
we compute the corresponding transformation property of the
dilatonic scalars, using the procedure illustrated in
(\ref{htrans}), we find
\begin{eqnarray}
\sigma_{v\neq u}^\prime&=&\sigma_{v\neq
u}\,\,;\,\,\,\sigma_{u}^\prime=-\sigma_u\,\,\Leftrightarrow\,\,\,
R_{v\neq u}^\prime=R_{v\neq
u}\,\,;\,\,\,\,R_u^\prime=\frac{1}{R_u}\,,\nonumber\\
\phi^\prime &=&\phi-\sigma_u\,,
\end{eqnarray}
which is the known effect of a $T$ duality along a single
direction. We can verify that under the effect of $T^{(w)}$ the
weight of $H_{uvw}$ is mapped into the weight of $T_{uv}{}^w$ and
moreover subsequent actions of $T^{(v)}$ and $T^{(u)}$ allow to
define the weights $W^Q{}_u{}^{vw}$ and $W^{R\,uvw}$, associated
with the non-geometric fluxes $Q_u{}^{vw}$ and $R^{uvw}$ in eq.
(\ref{htqr}) respectively
\begin{eqnarray}
W^H_{uvw}&\stackrel{T^{(w)}}{\longrightarrow }&
W^T_{uv}{}^{w}\stackrel{T^{(v)}}{\longrightarrow
}W^Q{}_u{}^{vw}\stackrel{T^{(u)}}{\longrightarrow
}\,\,W^{R\,uvw}\,\nonumber\\
W^Q{}_u{}^{vw}&=&\epsilon_u-\epsilon_v-\epsilon_w+\frac{1}{\sqrt{2}}\,\epsilon_{10}\,\,\,;\,\,\,\,\,
W^{R\,uvw}=-\epsilon_u-\epsilon_v-\epsilon_w+\frac{1}{\sqrt{2}}\,\epsilon_{10}\,.
\end{eqnarray}
The non perturbative $S$--duality is implemented as a Weyl
transformation with respect to the vector $\Delta=a$ in eq.
(\ref{hvect}), which is an $\mathfrak{e}_{7(7)}$ root in Type IIB
theory, but not in Type IIA theory, see equation (\ref{iiaiib}).
This represents the known fact that $S$--duality is a symmetry of
Type IIB theory (it corresponds to an ${\rm E}_{7(7)}$
transformation) but not of Type IIA theory (it maps Type IIA
superstring into M-theory). If we compute its action on the
dilatonic scalars, using (\ref{htrans}), we find
\begin{eqnarray}
\hat{\sigma}_u^\prime &=
&\hat{\sigma}_u\,\,\,;\,\,\,\,\phi^\prime=-\phi\,,
\end{eqnarray}
where $\hat{\sigma}_u$ are the radial moduli in the ten
dimensional Einstein frame. One can verify, using the weight
representation in eq. (\ref{weightflux}), that in Type IIB theory
\begin{eqnarray}
\sigma_a(W^H_{u_1 u_2 u_3})&=&W^F_{u_1 u_2 u_3}\,,
\end{eqnarray}
which is the known $S$--duality correspondence between the NS-NS
and the R-R 3--form fluxes $H^{(3)}_{u_1 u_2 u_3}$, $F^{(3)}_{u_1
u_2 u_3}$. One can also verify that the torsion $T_{uv}{}^w$ is
inert under $S$--duality, while the action of $S$--duality on the
non--geometric fluxes gives rise to more general fluxes which we
can identify with components of the embedding tensor, knowing the
corresponding weights. Finally let us consider the $U$--duality
transformation introduced at the end of section 7 to describe the
relation between ${\bf 20}_+$ and ${\bf 20}_-$ and between the
${\bf 1}_+$ and the ${\bf 1}_-$ representations in the branching
of the ${\bf 78}$ of ${\rm E}_{6(6)}$ with respect to ${\rm
SL}(6,\mathbb{R})\times {\rm SL}(2,\mathbb{R})$. The embedding of
${\rm E}_{6(6)}$ inside ${\rm E}_{7(7)}$ is defined by identifying
the simple roots of the former with $\alpha_2,\dots,\alpha_7$.
This duality transformation is implemented by the Weyl
transformation with respect to the root $\alpha$ of ${\rm
SL}(2,\mathbb{R})$, which has the form
\begin{eqnarray}
\alpha &=& a+\epsilon_5+\dots+\epsilon_9\,.
\end{eqnarray}
From the transformation property of $\vec{h}$ under
$\sigma_\alpha$ in (\ref{htrans}) we deduce the transformation
rules for the dilatonic scalars in eqs. (\ref{uduatra}).

\bibliographystyle{amsplain}

\begin{thebibliography}{}

  \bibitem{g}M.~Grana,
  {\it Flux compactifications in string theory: A comprehensive review},
  Phys.\ Rept.\  {\bf 423} (2006) 91;
 R.~Blumenhagen, B.~Kors, D.~Lust and S.~Stieberger,
 {\it Four-dimensional string compactifications with D-branes, orientifolds and
  fluxes},
  arXiv:hep-th/0610327;M.~R.~Douglas and S.~Kachru,
  {\it Flux compactification},
  arXiv:hep-th/0610102.
  \bibitem{formflux}
 J.~P.~Derendinger, L.~E.~Ibanez and H.~P.~Nilles,
  {\it On The Low-Energy D = 4, N=1 Supergravity Theory Extracted From The D = 10,
  N=1 Superstring},
  Phys.\ Lett.\ B {\bf 155} (1985) 65.
J.~Polchinski and A.~Strominger,
  {\it New Vacua for Type II String Theory},
  Phys.\ Lett.\ B {\bf 388} (1996) 736;
    K.~Dasgupta, G.~Rajesh and S.~Sethi,
  {\it M theory, orientifolds and G-flux},
  JHEP {\bf 9908} (1999) 023;
S.~Gukov, C.~Vafa and E.~Witten,
  {\it CFT's from Calabi-Yau four-folds},
  Nucl.\ Phys.\ B {\bf 584} (2000) 69
  [Erratum-ibid.\ B {\bf 608} (2001) 477].
   S.~B.~Giddings, S.~Kachru and J.~Polchinski,
  {\it Hierarchies from fluxes in string compactifications},
  Phys.\ Rev.\ D {\bf 66} (2002) 106006;
 G.~Dall'Agata,
  {\it Type IIB supergravity compactified on a Calabi-Yau manifold with
  H-fluxes},
  JHEP {\bf 0111} (2001) 005;
  A.~R.~Frey and J.~Polchinski,
 {\it N = 3 warped compactifications},
  Phys.\ Rev.\ D {\bf 65} (2002) 126009;
   S.~Kachru, M.~B.~Schulz and S.~Trivedi,
  {\it Moduli stabilization from fluxes in a simple IIB orientifold},
  JHEP {\bf 0310} (2003) 007;
 S.~Kachru, M.~B.~Schulz, P.~K.~Tripathy and S.~P.~Trivedi,
  {\it New supersymmetric string compactifications},
  JHEP {\bf 0303} (2003) 061;
 P.~K.~Tripathy and S.~P.~Trivedi,
  {\it Compactification with flux on K3 and tori},
  JHEP {\bf 0303} (2003) 028;
 R.~D'Auria, S.~Ferrara and S.~Vaula,
  {\it N = 4 gauged supergravity and a IIB orientifold with fluxes},
  New J.\ Phys.\  {\bf 4} (2002) 71;
 R.~D'Auria, S.~Ferrara, M.~A.~Lledo and S.~Vaula,
  {\it No-scale N = 4 supergravity coupled to Yang-Mills: The scalar potential and
  super Higgs effect},
  Phys.\ Lett.\ B {\bf 557} (2003) 278.
 S.~Kachru, R.~Kallosh, A.~Linde and S.~P.~Trivedi,
  {\it De Sitter vacua in string theory},
  Phys.\ Rev.\ D {\bf 68} (2003) 046005
  [arXiv:hep-th/0301240].
\bibitem{gargiulo} R.~D'Auria, S.~Ferrara, F.~Gargiulo, M.~Trigiante and S.~Vaula,
  {\it N = 4 supergravity Lagrangian for type IIB on T**6/Z(2) in presence of
  fluxes and D3-branes},
  JHEP {\bf 0306} (2003) 045;
M.~Berg, M.~Haack and B.~Kors,
  {\it An orientifold with fluxes and branes via T-duality},
  Nucl.\ Phys.\  B {\bf 669} (2003) 3.
\bibitem{formflux2}
  C.~Angelantonj, S.~Ferrara and M.~Trigiante,
 {\it New D = 4 gauged supergravities from N = 4 orientifolds with fluxes},
  JHEP {\bf 0310} (2003) 015;
C.~Angelantonj, S.~Ferrara and M.~Trigiante,
  {\it Unusual gauged supergravities from type IIA and type IIB orientifolds},
  Phys.\ Lett.\ B {\bf 582} (2004) 263;
L.~Andrianopoli, R.~D'Auria, S.~Ferrara and M.~A.~Lledo,
  JHEP {\bf 0303} (2003) 044;
 C.~Angelantonj, R.~D'Auria, S.~Ferrara and M.~Trigiante,
  {\it K3 x T**2/Z(2) orientifolds with fluxes, open string moduli and  critical
  points},
  Phys.\ Lett.\ B {\bf 583} (2004) 331;
 D.~Lust, S.~Reffert and S.~Stieberger,
  {\it Flux-induced soft supersymmetry breaking in chiral type IIb  orientifolds
  with D3/D7-branes},
  Nucl.\ Phys.\ B {\bf 706} (2005) 3;
 D.~Lust, S.~Reffert, W.~Schulgin and S.~Stieberger,
  {\it Moduli stabilization in type IIB orientifolds. I: Orbifold limits}, arXiv:hep-th/0506090;
D.~Lust, S.~Reffert, E.~Scheidegger, W.~Schulgin and S.~Stieberger,
  {\it Moduli stabilization in type IIB orientifolds. II}, arXiv:hep-th/0609013.
\bibitem{ss} J.~Scherk and J.~H.~Schwarz,
  {\it How To Get Masses From Extra Dimensions},
  Nucl.\ Phys.\ B {\bf 153} (1979) 61.
\bibitem{km} N.~Kaloper and R.~C.~Myers,
  {\it The O(dd) story of massive supergravity},
  JHEP {\bf 9905} (1999) 010.

\bibitem{ccdlmz}  G.~Lopes Cardoso, G.~Curio, G.~Dall'Agata, D.~Lust, P.~Manousselis and G.~Zoupanos,
  {\it Non-Kaehler string backgrounds and their five torsion classes},
  Nucl.\ Phys.\ B {\bf 652} (2003) 5;
M.~Grana, R.~Minasian, M.~Petrini and A.~Tomasiello,
  {\it Supersymmetric backgrounds from generalized Calabi-Yau manifolds},
  JHEP {\bf 0408} (2004) 046;
M.~Grana, J.~Louis and D.~Waldram,
  {\it Hitchin functionals in N = 2 supergravity},
  JHEP {\bf 0601} (2006) 008.
\bibitem{df}
 G.~Dall'Agata and S.~Ferrara,
  {\it Gauged supergravity algebras from twisted tori compactifications with
  fluxes},
  Nucl.\ Phys.\ B {\bf 717} (2005) 223.
\bibitem{alt} L.~Andrianopoli, M.~A.~Lledo and M.~Trigiante,
 {\it The Scherk-Schwarz mechanism as a flux compactification with internal
  torsion},
  JHEP {\bf 0505} (2005) 051.
  \bibitem{dft1}
 R.~D'Auria, S.~Ferrara and M.~Trigiante,
  {\it E(7)(7) symmetry and dual gauge algebra of M-theory on a twisted
  seven-torus},
  Nucl.\ Phys.\ B {\bf 732} (2006) 389.
  \bibitem{dft2}
   R.~D'Auria, S.~Ferrara and M.~Trigiante,
  {\it Curvatures and potential of M-theory in D = 4 with fluxes and twist},
  JHEP {\bf 0509} (2005) 035.
  \bibitem{dp} G.~Dall'Agata and N.~Prezas,
  {\it Scherk-Schwarz reduction of M-theory on G2-manifolds with fluxes},
  JHEP {\bf 0510} (2005) 103.
\bibitem{ft} P.~Fre' and M.~Trigiante,
  {\it Twisted tori and fluxes: A no go theorem for Lie groups of weak G(2)
  holonomy},
  Nucl.\ Phys.\ B {\bf 751} (2006) 343.
\bibitem{dkpz} J.~P.~Derendinger, C.~Kounnas, P.~M.~Petropoulos and F.~Zwirner,
 {\it Superpotentials in IIA compactifications with general fluxes},
  Nucl.\ Phys.\ B {\bf 715} (2005) 211.
  \bibitem{twistcosmo}I.~P.~Neupane and D.~L.~Wiltshire,
  {\it Cosmic acceleration from M theory on twisted spaces},
  Phys.\ Rev.\  D {\bf 72} (2005) 083509;
   I.~P.~Neupane,
  {\it Accelerating universes from compactification on a warped conifold},
  Phys.\ Rev.\ Lett.\  {\bf 98} (2007) 061301
  [arXiv:hep-th/0609086].
\bibitem{hre} C.~M.~Hull and R.~A.~Reid-Edwards,
  {\it Flux compactifications of M-theory on twisted tori},
  JHEP {\bf 0610} (2006) 086.
\bibitem{vz}G.~Villadoro and F.~Zwirner,
  {\it The minimal N = 4 no-scale model from generalized dimensional  reduction},
  JHEP {\bf 0407} (2004) 055;
G.~Villadoro and F.~Zwirner,
  {\it N = 1 effective potential from dual type-IIA D6/O6 orientifolds with
  general fluxes},
  JHEP {\bf 0506} (2005) 047.
\bibitem{mr} V.~Mathai and J.~M.~Rosenberg,
  {\it T-duality for torus bundles via noncommutative topology},
  Commun.\ Math.\ Phys.\  {\bf 253} (2004) 705.
\bibitem{h} C.~M.~Hull,
  {\it A geometry for non-geometric string backgrounds},
  JHEP {\bf 0510} (2005) 065.
\bibitem{dh} A.~Dabholkar and C.~Hull,
  {\it Duality twists, orbifolds, and fluxes},
  JHEP {\bf 0309} (2003) 054;
  {\it Generalised T-duality and non-geometric backgrounds},
  JHEP {\bf 0605} (2006) 009.
\bibitem{stw}J.~Shelton, W.~Taylor and B.~Wecht,
  {\it Nongeometric flux compactifications},
  JHEP {\bf 0510} (2005) 085.
\bibitem{glw2} M.~Grana, J.~Louis and D.~Waldram,
  {\it SU(3) x SU(3) compactification and mirror duals of magnetic fluxes},
  arXiv:hep-th/0612237.
\bibitem{dwn} B.~de Wit and H.~Nicolai,
  {\it N=8 Supergravity},
  Nucl.\ Phys.\ B {\bf 208} (1982) 323;
 C.~M.~Hull,
  {\it More Gaugings Of N=8 Supergravity},
  Phys.\ Lett.\ B {\bf 148} (1984) 297.
\bibitem{gaug} L.~Andrianopoli, M.~Bertolini, A.~Ceresole, R.~D'Auria, S.~Ferrara, P.~Fre and T.~Magri,
  {\it N = 2 supergravity and N = 2 super Yang-Mills theory on general scalar
  manifolds: Symplectic covariance, gaugings and the momentum map},
  J.\ Geom.\ Phys.\  {\bf 23} (1997) 111;
  M.~de Roo and P.~Wagemans,
  {\it Gauge Matter Coupling In N=4 Supergravity},
  Nucl.\ Phys.\ B {\bf 262} (1985) 644;
   J.~Schon and M.~Weidner,
  {\it Gauged N = 4 supergravities},
  JHEP {\bf 0605} (2006) 034.
\bibitem{cgftt} F.~Cordaro, P.~Fre, L.~Gualtieri, P.~Termonia and M.~Trigiante,
  {\it N = 8 gaugings revisited: An exhaustive classification},
  Nucl.\ Phys.\ B {\bf 532} (1998) 245.
\bibitem{adfl} L.~Andrianopoli, R.~D'Auria, S.~Ferrara and M.~A.~Lledo,
  {\it Gauging of flat groups in four dimensional supergravity},
  JHEP {\bf 0207} (2002) 010.
\bibitem{dwst1}B.~de Wit, H.~Samtleben and M.~Trigiante,
  {\it On Lagrangians and gaugings of maximal supergravities},
  Nucl.\ Phys.\ B {\bf 655} (2003) 93.
\bibitem{dwst2}  B.~de Wit, H.~Samtleben and M.~Trigiante,
  {\it Maximal supergravity from IIB flux compactifications},
  Phys.\ Lett.\ B {\bf 583} (2004) 338.
\bibitem{dws} B.~de Wit and H.~Samtleben,
  {\it Gauged maximal supergravities and hierarchies of nonabelian vector-tensor
  systems},
  Fortsch.\ Phys.\  {\bf 53} (2005) 442.
\bibitem{dwst3}
  B.~de Wit, H.~Samtleben and M.~Trigiante,
  {\it Magnetic charges in local field theory},
  JHEP {\bf 0509} (2005) 016.
\bibitem{noscale}E.~Cremmer, S.~Ferrara, C.~Kounnas and D.~V.~Nanopoulos,
Phys.\ Lett. {\bf B133}, 61 (1983);
J.~R.~Ellis, C.~Kounnas and D.~V.~Nanopoulos,
{\it No Scale Supersymmetric Guts},
 Nucl.\ Phys. {\bf B247}, 373 (1984);
R.~Barbieri, E.~Cremmer and S.~Ferrara,
{\it Flat And Positive Potentials In N=1 Supergravity},
Phys.\ Lett.\ B {\bf 163}, 143 (1985).
\bibitem{css} E.~Cremmer, J.~Scherk and J.~H.~Schwarz,
  {\it Spontaneously Broken N=8 Supergravity},
  Phys.\ Lett.\ B {\bf 84} (1979) 83.
\bibitem{ht} C.~M.~Hull and P.~K.~Townsend,
  {\it Unity of superstring dualities},
  Nucl.\ Phys.\ B {\bf 438} (1995) 109.
\bibitem{dd}
 E.~Cremmer, B.~Julia, H.~Lu and C.~N.~Pope,
 {\it Dualisation of dualities. I},
  Nucl.\ Phys.\ B {\bf 523} (1998) 73;
 E.~Cremmer, B.~Julia, H.~Lu and C.~N.~Pope,
  {\it Dualisation of dualities. II: Twisted self-duality of doubled fields  and
  superdualities},
  Nucl.\ Phys.\ B {\bf 535} (1998) 242.
\bibitem{cj}
  E.~Cremmer and B.~Julia,
  {\it The SO(8) Supergravity},
  Nucl.\ Phys.\ B {\bf 159} (1979) 141.
\bibitem{cjs}  E.~Cremmer, B.~Julia and J.~Scherk,
  {\it Supergravity theory in 11 dimensions},
  Phys.\ Lett.\ B {\bf 76} (1978) 409.
\bibitem{gz}  M.~K.~Gaillard and B.~Zumino,
  {\it Duality Rotations For Interacting Fields},
  Nucl.\ Phys.\ B {\bf 193} (1981) 221.
\bibitem{dwlv}B.~de Wit, P.~G.~Lauwers and A.~Van Proeyen,
  {\it Lagrangians Of N=2 Supergravity - Matter Systems},
  Nucl.\ Phys.\ B {\bf 255} (1985) 569.
\bibitem{previous} D.~Z.~Freedman and P.~K.~Townsend,
  {\it Antisymmetric Tensor Gauge Theories And Nonlinear Sigma Models},
  Nucl.\ Phys.\ B {\bf 177} (1981) 282;
 M.~Henneaux and B.~Knaepen,
  {\it All consistent interactions for exterior form gauge fields},
  Phys.\ Rev.\ D {\bf 56} (1997) 6076;
  R.~D'Auria, L.~Sommovigo and S.~Vaula,
  {\it N = 2 supergravity Lagrangian coupled to tensor multiplets with  electric
  and magnetic fluxes},
  JHEP {\bf 0411} (2004) 028
\bibitem{s} L.~Sommovigo,
  {\it Poincare dual of D = 4 N = 2 supergravity with tensor multiplets},
  Nucl.\ Phys.\ B {\bf 716} (2005) 248.
\bibitem{pope}H.~Lu, C.~N.~Pope and K.~S.~Stelle,
  {\it Weyl Group Invariance and p-brane Multiplets},
  Nucl.\ Phys.\ B {\bf 476} (1996) 89.
  \bibitem{micu}
  I.~Benmachiche and T.~W.~Grimm,
  {\it Generalized N = 1 orientifold compactifications and the Hitchin
  functionals},
  Nucl.\ Phys.\  B {\bf 748} (2006) 200;
  A.~Micu, E.~Palti and G.~Tasinato,
 {\it Towards Minkowski Vacua in Type II String Compactifications},
  arXiv:hep-th/0701173.
\bibitem{heis} R.~D'Auria, S.~Ferrara, M.~Trigiante and S.~Vaula,
  {\it Gauging the Heisenberg algebra of special quaternionic manifolds},
  Phys.\ Lett.\ B {\bf 610} (2005) 147;
R.~D'Auria, S.~Ferrara, M.~Trigiante and S.~Vaula,
  {\it Scalar potential for the gauged Heisenberg algebra and a non-polynomial
  antisymmetric tensor theory},
  Phys.\ Lett.\ B {\bf 610} (2005) 270.
  \bibitem{heis2}R.~D'Auria, S.~Ferrara and M.~Trigiante,
  {\it On the supergravity formulation of mirror symmetry in generalized
  Calabi-Yau manifolds},
  arXiv:hep-th/0701247.
\bibitem{sla} L.~Andrianopoli, R.~D'Auria, S.~Ferrara, P.~Fre and M.~Trigiante,
 {\it R-R scalars, U-duality and solvable Lie algebras},
  Nucl.\ Phys.\ B {\bf 496} (1997) 617;
   L.~Andrianopoli, R.~D'Auria, S.~Ferrara, P.~Fre, R.~Minasian and M.~Trigiante,
  {\it Solvable Lie algebras in type IIA, type IIB and M theories},
  Nucl.\ Phys.\ B {\bf 493} (1997) 249
  [arXiv:hep-th/9612202].
  \bibitem{RRbh} M.~Bertolini and M.~Trigiante,
 {\it Regular R-R and NS-NS BPS black holes},
  Int.\ J.\ Mod.\ Phys.\ A {\bf 15} (2000) 5017.
\end{thebibliography}
 
\end{document}